%% file: main.tex
\documentclass[journal]{vgtc}                
\ifpdf
  \pdfoutput=1\relax                   
  \usepackage{graphicx}                
  \DeclareGraphicsExtensions{.pdf,.png,.jpg,.jpeg} 
\else
  \usepackage{graphicx}                
  \DeclareGraphicsExtensions{.eps}     
\fi%

\graphicspath{{figures/}{pictures/}{images/}{./}} 

\usepackage{microtype}                 
\PassOptionsToPackage{warn}{textcomp}  
\usepackage{textcomp}                  
\usepackage{mathptmx}                  
\usepackage{times}                     
\usepackage{cite}                      
\usepackage{tabu}                      
\usepackage{booktabs}                  

\usepackage{balance}  
\usepackage{graphics} 
\usepackage{times}    
\usepackage{url}      
\usepackage[ruled,vlined,linesnumbered]{algorithm2e}
\usepackage{enumitem}
\usepackage{xspace}
\usepackage[table,usenames,dvipsnames]{xcolor}
\usepackage{booktabs}
\usepackage{multirow}
\usepackage{array}
\usepackage{cellspace}
\usepackage[10pt]{moresize} 
\usepackage{amsmath}
\usepackage{listings}
\usepackage{xcolor}
\usepackage{amsmath}
\usepackage{MnSymbol}

\usepackage{circledsteps}

\onlineid{1495}

\vgtccategory{Research}
\vgtcpapertype{Representations and Interaction}

\newcommand{\name}{Erato\xspace}
\newcommand{\etal}{{\it et~al.}\xspace}
\newcommand{\type}{\textit{\textbf{type}}\xspace}
\newcommand{\subspace}{\textit{\textbf{subspace}}\xspace}
\newcommand{\measure}{\textit{\textbf{measure}}\xspace}
\newcommand{\focus}{\textit{\textbf{focus}}\xspace}
\newcommand{\breakdown}{\textit{\textbf{breakdown}}\xspace}

\newcommand{\factsheet}{\textit{\textbf{factsheet}}\xspace}

\newcommand{\cn}[1]{\textcolor{black}{#1}}
\newcommand{\smd}[1]{\textcolor{black}{#1}}
\newcommand{\menndy}[1]{\textcolor{black}{#1}}

\title{\name: Cooperative Data Story Editing via Fact Interpolation}

\author{Mengdi Sun, Ligan Cai, Weiwei Cui, Yanqiu Wu, Yang Shi, and Nan Cao}
\authorfooter{
\item
 Mengdi Sun, Ligan Cai, Yanqiu Wu, Yang Shi, and Nan Cao are with Intelligent Big Data Visualization Lab at Tongji University.\\E-mails:\{menndy, tsailgan\}@tongji.edu.cn, \{wuyanqiu.idvx, shiyang1230, nan.cao\}@gmail.com. Nan Cao is the corresponding author.
\item
 Weiwei Cui is with Microsoft Research Asia. \\E-mail: weiwei.cui@microsoft.com.
}

\shortauthortitle{Biv \MakeLowercase{\textit{et al.}}: Global Illumination for Fun and Profit}

\abstract{
As an effective form of narrative visualization, visual data stories are widely used in data-driven storytelling to communicate complex insights and support data understanding.
Although important, they are difficult to create, as a variety of interdisciplinary skills, such as data analysis and design, are required.
In this work, we introduce \name, a human-machine cooperative data story editing system, which allows users to generate insightful and fluent data stories together with the computer.
\cn{Specifically, \name \menndy{only requires a number of keyframes provided by the user to briefly describe the topic and structure of a data story.} Meanwhile, our system leverages a novel interpolation algorithm to help users insert intermediate frames between the keyframes to smooth the transition}.
We evaluated the effectiveness and usefulness of the \name system via a series of evaluations including a Turing test, a controlled user study, a performance validation, and interviews with three expert users. The evaluation results showed that the proposed interpolation technique was able to generate coherent story content and help users create data stories more efficiently.

} 


\CCScatlist{ 
 \CCScat{K.6.1}{Management of Computing and Information Systems}%
{Project and People Management}{Life Cycle};
 \CCScat{K.7.m}{The Computing Profession}{Miscellaneous}{Ethics}
}

\teaser{
  \centering
  \includegraphics[width=\linewidth]{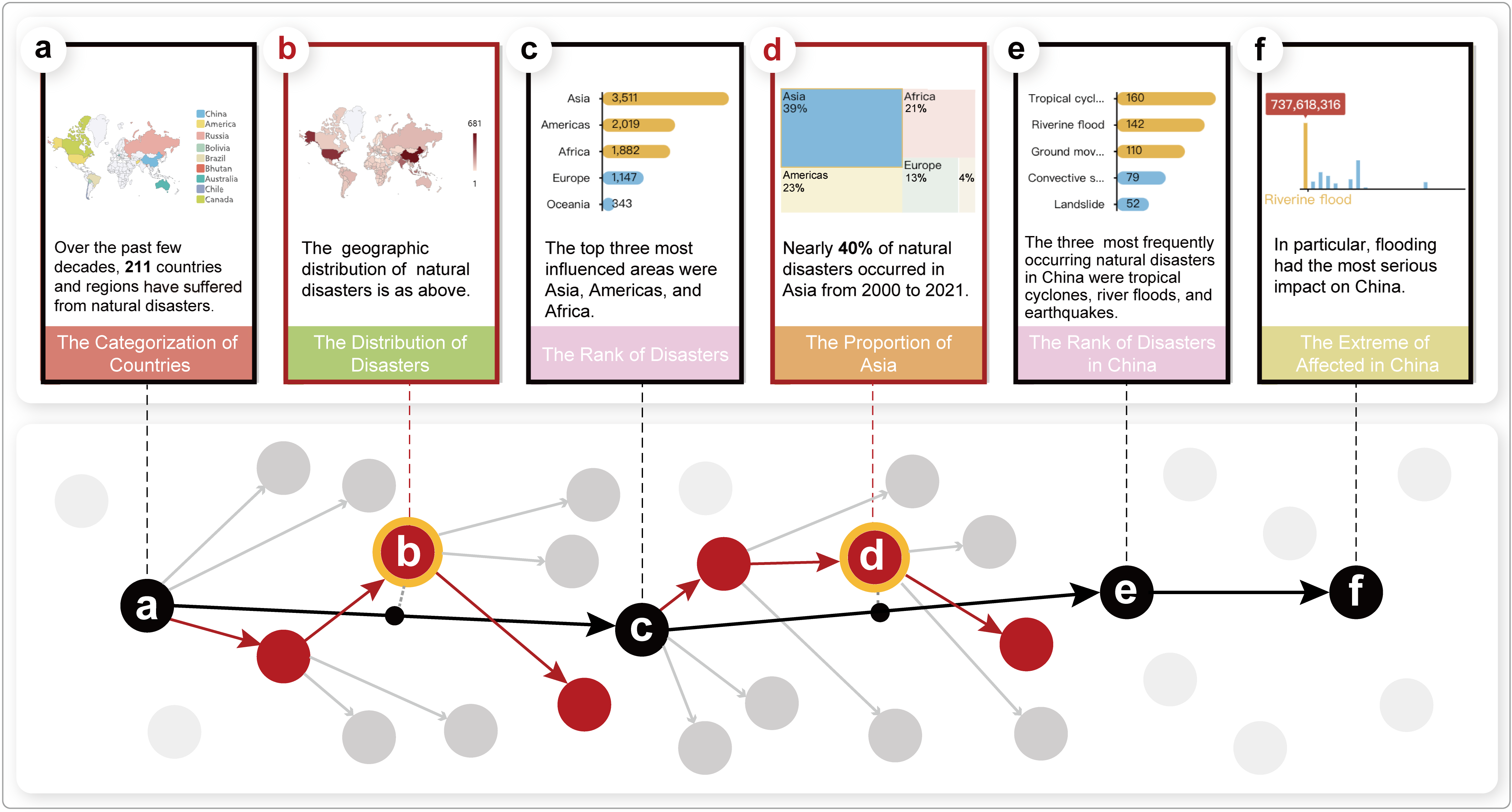}
  \caption{
  \menndy{A data story about natural disasters represented in form of a storyline that was authored by a professional data analyst.} The data facts in black (a, c, e, f) were created by the analyst as keyframes of the story, while the facts in red (b, d) were generated based on our interpolation algorithm. The algorithm searches through the fact space to find data facts that best fill the content gap between two keyframes. The story first illustrates an overall situation of global natural disasters (a-c) and gradually focuses on the situation in China (d,e). Finally, it reveals that floods have the most pernicious impact on China (f).
  The corresponding interpolation process is also shown under the data story. The searching path is marked in red and the nodes with yellow borders are the final selected interpolation results.}
  \label{fig:teaser}
}

\vgtcinsertpkg

\begin{document}



\input{sections/01-intro.tex}

\input{sections/02-related}
\input{sections/03-system}
\input{sections/04-technique}
\input{sections/05-interface}
\input{sections/06-evaluation}
\input{sections/07-conclusion}

\acknowledgments{
Nan Cao is the corresponding author. This work was supported by NSFC 62072338, 62061136003 and NSF Shanghai 20ZR1461500.}

\bibliographystyle{abbrv-doi}

\bibliography{main}
\end{document}

%% file: sections/01-intro.tex
\firstsection{Introduction}
\maketitle

A visual data story is a series of connected data facts shown in form of a narrative visualization, which is usually used to help with information communication~\cite{lee2015more}. It has been widely used in many application domains such as business intelligence, data journalism, advertising, and education~\cite{segel2010narrative}. Although important, creating a data story is not an easy task as it acquires multiple skills including data analysis, visualization, graphic design, and storytelling.

To facilitate the creation of data stories, during the past decades, theories, techniques, and tools have been extensively studied and developed. For example, a series of design spaces have been proposed from two major aspects: the narrative structures~\cite{yang2021design,lan2021understanding} and the visual representation methods~\cite{shi2021communicating,lan2021kineticharts,shi2021understanding,hullman2013deeper,mckenna2017visual,segel2010narrative,stolper2016emerging}, while taking communication goals and tasks into consideration. At the same time, to lower the technical barriers, a number of interactive authoring tools that integrate advanced data analysis and visualization functionalities have been developed~\cite{chen2018supporting,liu2018data,wang2018infonice,wang_narvis_2019}. Although very helpful, using design space or authoring tools to design and create a data story majorly rely on users to make decisions and take actions. The man-made stories are insightful and fluid, but the process is cumbersome, tedious, and inefficient. To address this issue, recent studies have been focusing on automatic data story generation based on intelligent algorithms~\cite{shi_task-oriented_2019,shi_calliope_2021,shi2021autoclips,zhao_chartstory_2021,wang_datashot_2020}. These techniques, although efficient, suffer from poor quality of the generation results and the lack of humanity, which is usually considered the soul of a story. Therefore, there is a gap between manual story authoring and automatic story generation. A tool that supports human-machine collaborative data story design and editing is desired.

To fill the gap, we introduce \name, a human-machine collaborative data story editing system, through which users can design and generate a data story together with a computer. In particular, using the system, \cn{a user only needs to concentrate on the key message by inputting a few keyframes (i.e., key data facts)}. \cn{The system will efficiently generate more details of the story by interpolating between any of the two succeeding keyframes}. Users can edit both the keyframes and the generated intermediate data facts at any time. When the keyframes are changed, the story content will be updated accordingly. In this way, \cn{the system \menndy{supports} the human-machine collaboration to generate data stories efficiently while keeping its humanity via the \smd{human-generated} keyframes.} The major contributions of the paper are as follows:

\begin{itemize}[leftmargin=10pt,topsep=2pt]
\itemsep -.5mm
\item {\bf System.} We introduce the first intelligent system, to the best of our knowledge, which is designed to support human-machine cooperative data story design and editing\footnote{\url{https://erato.idvxlab.com/project/}}. Based on the system, a user can easily generate a data story by only editing a few keyframes and the system will fill the gap by generating a series of data facts to connect the succeeding keyframes. 

\item {\bf Data Fact Interpolation Technique.} We introduce the first interpolation technique that is able to linearly interpolate between two data facts (usually keyframes in a data story) to generate a series of meaningfully and smoothly connected data facts.

\item {\bf Evaluation.} We demonstrate the utility of the proposed system via an interview and case study with three expert users and also show the performance of the fact embedding model via a quantitative evaluation and a controlled user study. A Turing test is also performed to evaluate the overall quality of the story generated based on our technique.
\end{itemize}

%% file: sections/02-related.tex
\section{Related Work}
Our work draws inspiration from three areas, including data-driven storytelling, automatic data visualization, and interpolation techniques in data visualization.

\subsection{Data-Driven Storytelling}
Visual data stories refer to a set of \smd{story} pieces that are visually presented in a meaningful way to deliver an intended message~\cite{segel2010narrative, lee2015more,tong2018storytelling,riche2018data}. Studies showed that incorporating data visualizations in concert with narrative could reveal information effectively~\cite{gershon2001storytelling} and enhance readers’ engagement, memory, comprehension, and communication~\cite{segel2010narrative,borkin2015beyond,dove2012narrative}. Therefore, visual data storytelling has gained increasing popularity in many domains and evolved into an important topic in the visualization community~\cite{kosara2013storytelling}.

Due to its importance, many theoretical design spaces have been introduced to help clarify the key concepts about visual narratives~\cite{segel2010narrative,mckenna2017visual,stolper2016emerging} and provide fundamental design principles~\cite{yang2021design,lan2021understanding,shi2021communicating,lan2021kineticharts,shi2021understanding}. Early studies focused on high-level concepts. For example, Segel and Heer~\cite{segel2010narrative} classified narrative visualization into seven genres. McKenna~\etal~\cite{mckenna2017visual} identified seven key factors for building a fluent visual narrative flow.
Stolper~\etal~\cite{stolper2016emerging} presented four high-level categories of narrative visualization techniques. Recent studies introduce several design spaces that more elaborately provide fundamental design principles from two major aspects: the narrative structures~\cite{yang2021design,lan2021understanding} and the visual representation methods~\cite{shi2021communicating,lan2021kineticharts,shi2021understanding}, while taking communication goals and design tasks into consideration. 

Based on these theories, techniques and authoring tools have also been proposed to help with the design and creation of visual data stories.
For example, chart sequencing techniques~\cite{hullman2013deeper,kim2017graphscape,shi_task-oriented_2019}  have been extensively studied for generating a data story by connecting charts to form a meaningful sequence.
Some authoring tools~\cite{bryan2016temporal,satyanarayan_authoring_2014,ren2017chartaccent} aim to help users interactively create and place custom annotations to generate a visual narrative, while others are introduced to support data videos~\cite{amini_authoring_2017}, and time-oriented storytelling~\cite{fulda_timelinecurator_2016,brehmer_timeline_2019}.
\cn{Datatoon~\cite{kim2019datatoon} is similar to our system in terms of leveraging interpolation techniques. However, we target at a different type of data story in this work, which requires a totally different set of techniques.}

Although very helpful, using the above design space or authoring tools to design and create a data story majorly rely on users to make decisions and take actions. Sometimes, the process is cumbersome, tedious, and inefficient. To address this issue, recent studies have been focusing on automatic data story generation based on intelligent algorithms~\cite{chen2018supporting,shi_calliope_2021,shi2021autoclips,zhao_chartstory_2021}. 
For example, Chen~\etal~\cite{chen2018supporting} proposed a framework and automatic workflow to bridge the gap between data analysis and communication.
AutoClips~\cite{shi2021autoclips} optimally organizes data facts in a parallel structure to create data videos.
Calliope~\cite{shi_calliope_2021} automatically generates data stories from a spreadsheet. ChartStory~\cite{zhao_chartstory_2021} characterizes charts by their similarity in a fixed layout to form a data story.  Although efficient, these tools suffer from their generation quality and lack of user engagement.

Different from the aforementioned techniques, \cn{\name fills the gap between manually data story authoring and automatic data story generation by striking a balance between machine and human involvements. It ensures the quality of stories, enhances engagement, and improves the efficiency of authoring at the same time}.

\subsection{Automatic Data Visualization}
Our work is also related to the broader area of automatic data visualization. As a visual data story is usually composed of individual data visualizations, the automatic extraction and visualization of data insights are essential to the efficiency of the story generation. 

There have been various studies of automatic data visualization over the decades, which can be largely classified into two categories: rule-based techniques and machine learning-based techniques. 
Rule-based techniques often derive from experimental findings and expert experience. For example, Mackinlay’s APT~\cite{mackinlay_automating_1986} and Sage~\cite{roth1994interactive} use expressiveness and perceptual effectiveness criteria to enumerate, filter, and score visualizations.
Show Me~\cite{mackinlay_show_2007}, Voyager~\cite{wongsuphasawat_voyager_2016,wongsuphasawat_voyager_2017} and DIVE~\cite{hu_dive_2018} extend the above approach by checking the data types. The machine learning-based approaches train a model to recommend charts or visual encoding methods for input data. For example, Data2Vis~\cite{dibia_data2vis_2019} learns an end-to-end generation model to translate data into visualization.
DeepEye~\cite{luo_deepeye_2018} and Draco~\cite{moritz2018formalizing} train a supervised learning-to-rank model to recommend visualizations.
VizML~\cite{hu_vizml_2019} integrates the interpretable measures of feature importance into automatic visualization tools.
Text2Vis~\cite{cui2019text} employs a natural language processing model to analyze data entities and convert them into proportion charts. Draco~\cite{moritz2018formalizing} employs answer set programming to automatically find out visual design violations. VizLinter~\cite{chen2021vizlinter} makes a step further by automatically fixing these violations via linear programming. To ensure efficiency, \name borrows rule-based methods used in~\cite{shi_calliope_2021} to automatically visualize story pieces in a number of predefined visualization charts.

Some other studies capture potentially interesting visualizations based on the statistical properties and insights of the input data, also known as auto-insights. 
For example, Foresight~\cite{demiralp2017foresight} helps users rapidly discover visual insights from large high-dimensional datasets.
AutoVis~\cite{wills_autovis_2010} recommends interesting relationships between variables in the data.
Tang~\etal~\cite{tang_extracting_2017} and Vartak~\etal~\cite{vartak_seedb_2015} captured interesting observations derived from aggregation results. 
DataShot~\cite{wang_datashot_2020} randomly extracts important insights from the data via statistical methods and displays them in form of a factsheet.  Calliope~\cite{shi_calliope_2021} makes a step further by searching through the data space to extract informative and logically connected data insights to generate a visual data story.
Different from these techniques, \name generates and visualizes informative data insights via interpolating between keyframes in a data story.

\subsection{Interpolation Techniques}
In the field of data visualization, interpolation techniques have been frequently used in animated transitions to smooth out a visual difference. For example, Wittenbrink~\cite{wittenbrink_ifs_1995} proposed a fractal interpolation for two or three dimension visualization.
Schlegel~\etal~\cite{schlegel_interpolation_2012} incorporated Gaussian process regression to interpolate uncertain data.
\cn{Gemini~\cite{kim2020gemini,kim2021gemini} recommends animated transitions between charts based on graphical interpolation.} In addition, based on specific application scenarios, a variety of interpolation techniques, such as image interpolation~\cite{thevenaz2000image}, sequence interpolation~\cite{ueng2005interpolation}, and surface interpolation~\cite{sarfraz_visualization_2002}, have been successfully exploited. However, existing techniques interpolate between graphic elements or visual attributes, none of them 
are able to directly interpolate the data to generate meaningful content used for authoring a data story. In this paper, we introduce a novel interpolation algorithm that is able to interpolate between two data facts to generate the content of story pieces directly.

%% file: sections/03-system.tex
\section{System Overview}
\label{sec:overview}

This section introduces the design of \name system. We first provide a formal definition of a data story and the corresponding notations that are used throughout the paper. After that, we summarize the design requirements and introduce the architectural design of the system. 

\subsection{Data Story}
In this paper, we borrow and slightly simplify the data story definition introduced in~\cite{shi_calliope_2021}. In this section, we briefly introduce the key concepts and notations used in this paper as the background but leave the details in~\cite{shi_calliope_2021} for readers to reference. In particular, we define a data story as a sequence of meaningfully connected data facts that are ordered according to narrative logic. It can be formally represented as  $\{f_1, f_2, \cdots, f_n\}$, where $f_i$ is a data fact, the elementary building block of the data story. Each fact provides a piece of information extracted from the data, which is formally given by a 5-tuple:
\[
\begin{aligned}
    f_i &= \{type, subspace, breakdown, measure, focus\}\\
    &= \{t_i, s_i, b_i, m_i, x_i\}
\end{aligned}
\]
where \type (denoted as $t_i$) indicates the type of information described by the fact. Similar to Calliope, \name also supports 10 fact types, which are \textit{value}, \textit{difference}, \textit{proportion}, \textit{trend}, \textit{categorization}, \textit{distribution}, \textit{rank}, \textit{association}, \textit{extreme}, and \textit{outlier}; 
\subspace (denoted as $s_i$) is given by a set of data filters, i.e., $\{\{\mathcal{F}_1 = \mathcal{V}_1\},\cdots, \{\mathcal{F}_k = \mathcal{V}_k\}\}$, which restrict the data scope of the fact. $\mathcal{F}_i$ and $\mathcal{V}_i$ respectively indicate a data field and the selected field value. \breakdown (denote\smd{d} as $b_i$) is a temporal or categorical data field, which divides a subspace into groups. Each group can be further measured based on a numerical field, indicating by \measure (denoted as $m_i$) via one of the following aggregation methods: count, sum, average, minimum, or maximum. \focus (denoted as $x_i$) indicates a data item or a group that needs to pay attention. For example, regarding to the following fact about the Winter Olympic Games 2022 in Beijing: {\it \{``distribution", \{\{Sex =``Female"\}\}, \{Country\}, \{sum(Gold Medal)\}, \{Country=``China"\}\}}, it indicates ``the distribution (fact \type) of the total number of (aggregation method) the gold medals (\measure) won by females (\subspace) across all the countries (\breakdown) and China is the highlight (\focus)".

\subsection{Design of \name System}
The design of \name system was inspired by the users' feedback collected during a series of 5 workshops on data story design, which were organized either by ourselves or by our colleagues in the last year. These workshops involved a total number of 125 participants with various backgrounds, such as university students (major in design, journalism and communication, and computer science), data journalists, citizen journalists, We-Media operators, and data analysts from consultant/IT companies. The goal of the workshops was to teach participants how to author a data story. Each workshop has a focused topic such as ``narrative structure", ``visualization and infographic design", ``data video authoring", ``data insight discovery", and ``logic transition and animation design". The participants were asked to use design and data analysis tools such as Adobe Illustrator, After Effects, Tableau, and Calliope\footnote{\url{https://datacalliope.com}} (a tool developed for automatic data story generation). 

Through these workshops, we collected a large number of valuable feedback, which eventually inspired us to design and develop the cooperative authoring tool introduced in this paper. The feedback mainly focused on the authoring experience and resulting stories. For example, most of the users felt using these tools to create a data story needed a lot of operations, which was quite inefficient and required much design and data analysis background. They generally liked the idea of automatic data story generation, but they felt the quality of the stories generated by Calliope was not satisfactory. They also believed Calliope limited their \menndy{control and involvement}. We have summarized their comments as the following design requirements:

\vspace{-0.8em}
\begin{enumerate}
\itemsep -1mm
\item[{\bf R1}] {\bf Incorporating Users' \menndy{Control and Involvement}.} To ensure the quality of a data story and \menndy{incorporate users' ideas}, the system should directly let the users decide what to tell (i.e., the key message) and how to tell (i.e., the narrative structure) a data story. 

\item[{\bf R2}] {\bf Improving the Authoring Efficiency.} During the data story authoring process, the system should be able to automatically deal with the tedious and cumbersome operations such as exploring the vast data space for potential story pieces or permutably arranging the data facts into a narrative structure. 

\item[{\bf R3}] {\bf \cn{Supporting Smart Interactive Authoring Mechanism.}} \cn{The system should provide a \smd{flexible and interactive} mechanism that is smart enough to automatically finish some \smd{time-consuming tasks} to accelerate the authoring process.}
In this way, users could collaborate with the system toward the same goal of creating a data story.

\end{enumerate}



\begin{figure}[tb]
  \centering
  \includegraphics[width=\linewidth]{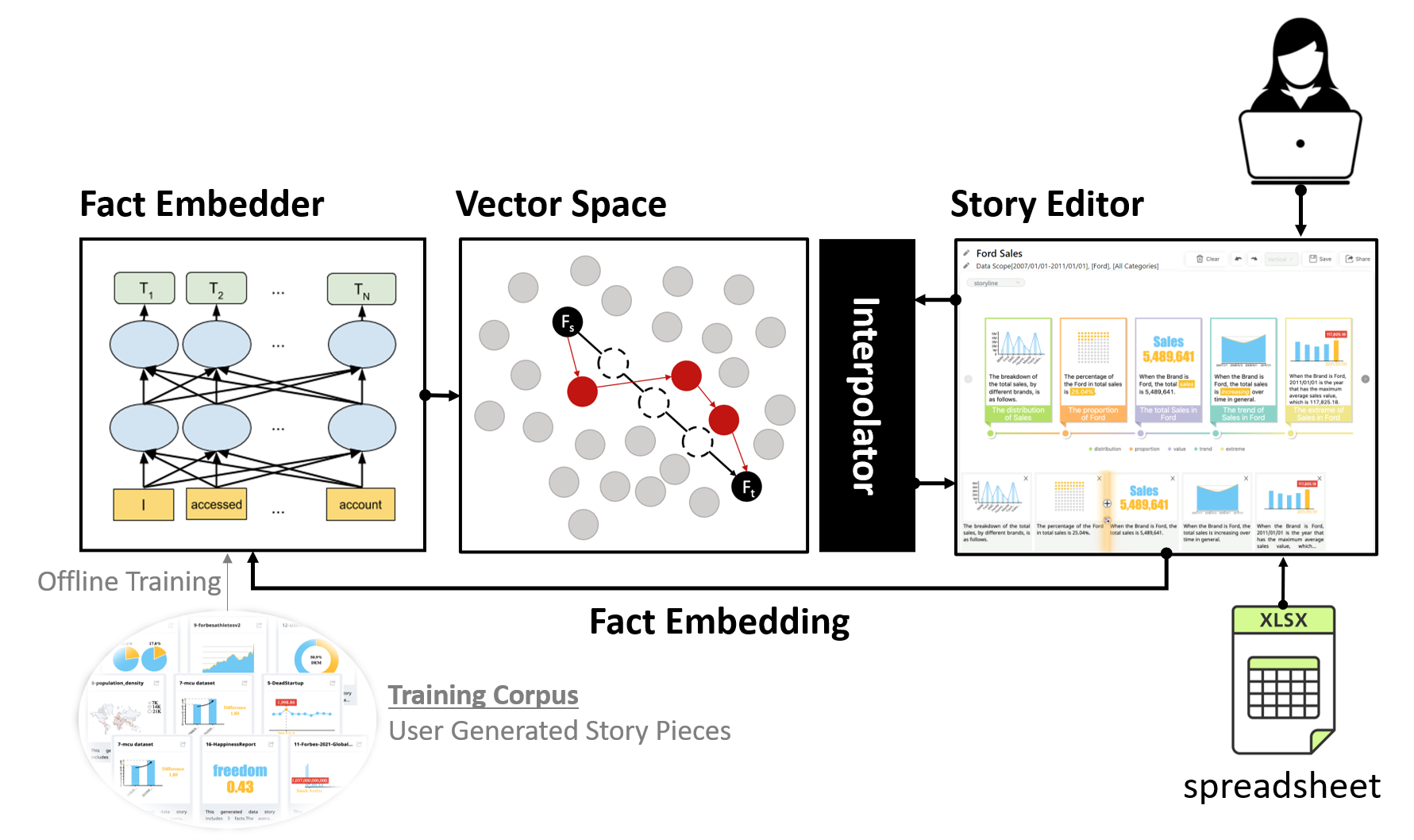}
  \vspace{-2em}
  \caption{The running pipeline of the \name system.} 
  \label{fig:pipeline}
  \vspace{-1.5em}
\end{figure}

\vspace{-0.8em}
To fulfill the requirements, we developed a cooperative data story authoring system, namely \name, based on a novel fact interpolation technique introduced in this paper. Fig.~\ref{fig:pipeline} illustrates the architecture and running pipeline of the system, which consists of three major modules, including the \textit{\underline{Fact Embedder}}, the  \textit{\underline{Interpolator}}, and the \textit{\underline{Story Editor}}. Generally, a user starts creating a data story by interactively inputting a number of data facts as key frames and arranging them into a storyline via the \textit{\underline{Story Editor}}. In this way, the user decides what to tell (keyframes) and how to tell (narrative structure) the story (\textbf{R1}). After that, the keyframes in the storyline are converted into corresponding vector representations and are projected into a vector space by the \textit{\underline{Fact Embedder}}. \smd{The} \textit{\underline{Fact Embedder}} is a pre-trained deep learning model that takes a fact's specification string as the input and converts it into a vector representation to facilitate numerical calculation. Finally, the \textit{\underline{Interpolator}} approximately interpolates between the vectors of two specified succeeding key frames to generate viable data facts by searching through the vector space (\textbf{R2}). The results are considered intermediate data facts between these two key frames and presented in the \textit{\underline{Story Editor}}.
Then the user can further verify, refine, and incorporate them to make a more smooth and more compelling story (\textbf{R3}).

%% file: sections/04-technique.tex
\input{tables/actions}
\vspace{-1.5em}
\section{Data Fact Interpolation}
In the system, a novel algorithm has been introduced to interpolate between data facts (i.e., key frames) and help users smooth transitions in a story. It first projects data facts to a vector space based on an embedding model. After that, for a selected pair of succeeding key frames, the algorithm linearly interpolates the corresponding vectors to generate a series of continuously changed latent vectors. Meanwhile, it searches through the fact space to find a series of data facts that best match with the latent vectors as the interpolation results.

\subsection{Fact Embedder}
\label{sec:model}

\begin{figure}[!t]
  \centering
  \vspace{-0.5em}
  \includegraphics[width=\linewidth]{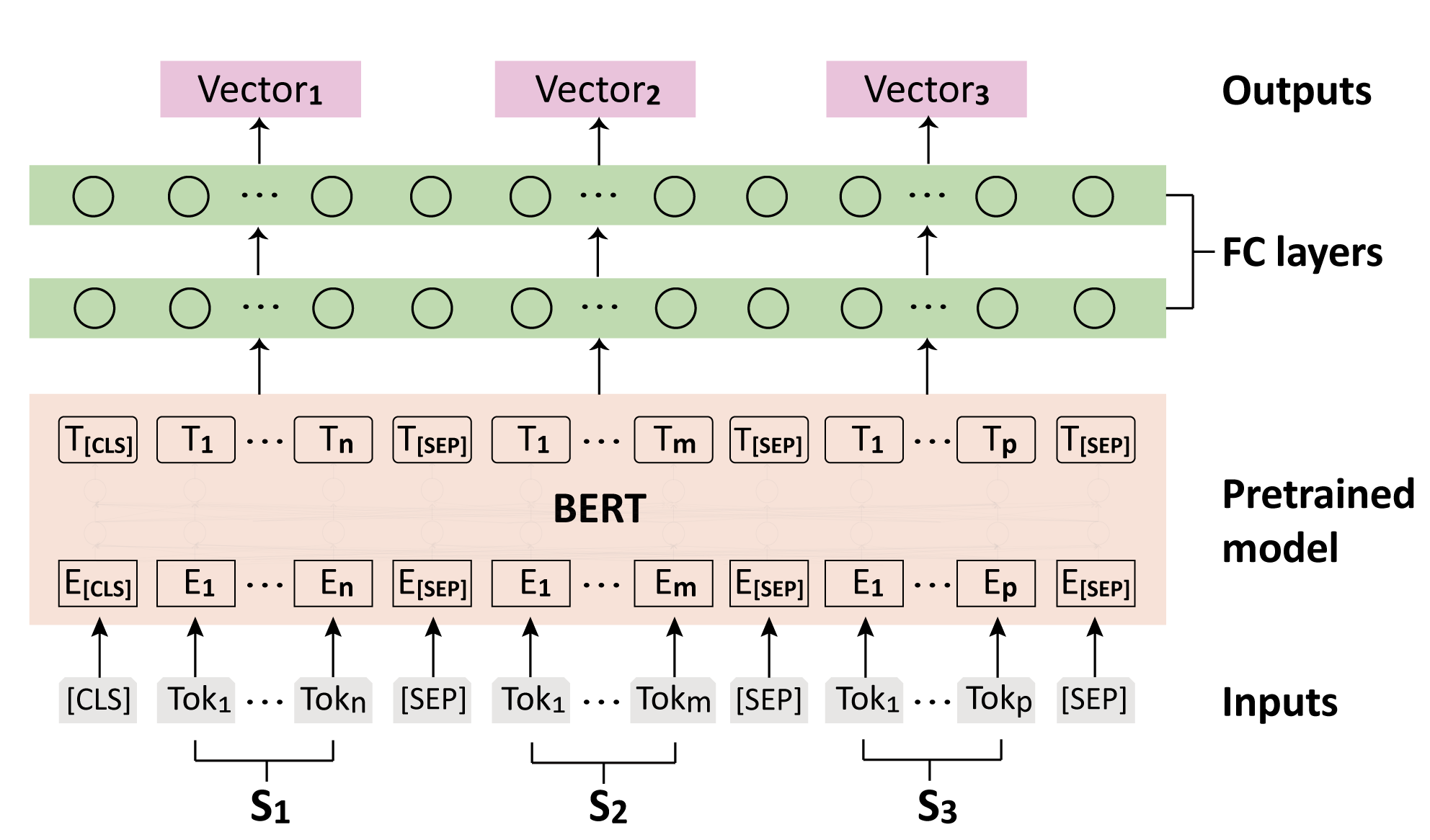}
  \vspace{-2em}
  \caption{Schematic diagrams of the fact embedding model.} 
  \label{fig:model}
  \vspace{-1.5em}
\end{figure}

The fact embedder employs a deep embedding model to covert data facts into vector representations. To support fact interpolation, the embedding should satisfy two criteria: ({\bf C1}) a vector should capture the data semantics of the corresponding fact; ({\bf C2}) interpolating between two vectors, $(v_s, v_t)$, should generate meaningful results that are not only numerically between $(v_s, v_t)$ but also semantically between the corresponding facts. 

With the above requirements in mind, we built a representation learning model (Fig.~\ref{fig:model}) for data facts based on BERT~\cite{devlin2018bert}, a pre-trained language representation model following the Transformer~\cite{vaswani2017attention} architecture. The major advantages of adopting BERT are three folds: (1) it adopts a masked mechanism and next sentence prediction to respectively provide a vector representation of both words and sentences that better captures the meaning of an input string; (2) it is trained based on a large text corpus, thus having a good generalization capability; (3) studies showed that BERT could be fine-tuned for a specific task based only on a small set of training samples~\cite{reimers-2019-sentence-bert}. 

Given all the above benefits, we implemented our representation learning model by directly adding two fully connected layers on top of BERT as shown in Fig.~\ref{fig:model}. Then, we fine-tuned the model based on a training set collected from a set of manually designed visual narratives, each of which essentially was a sequence of logically connected data facts. Specifically, each training value was a trigram of \smd{the} facts in the dataset, which captured logical relationships between facts but, at the same time, reduced the unnecessary complexity caused by long stories. We converted each of the data fact in a trigram into a tokenized string to fit the input format of our representation learning model: ``{\it [type]fact-type [subspace]field,value [measure]field,agg [breakdown]field [focus]value [meta]extra-info}". In addition to the five tuples for a fact (Section~\ref{sec:overview}), we added the \underline{\textit{meta}} information to provide extra useful information about the fact. For example, when the fact type is ``trend", the extra information will be ``increasing" or ``decreasing", which indicates the specific type of trend; when the fact type is ``extreme", the extra information will be ``minimum" or ``maximum". The experiment showed that adding this information will increase the accuracy of the embedding results. 

\begin{figure}[!t]
  \centering
  \vspace{-0.5em}
  \includegraphics[width=\linewidth]{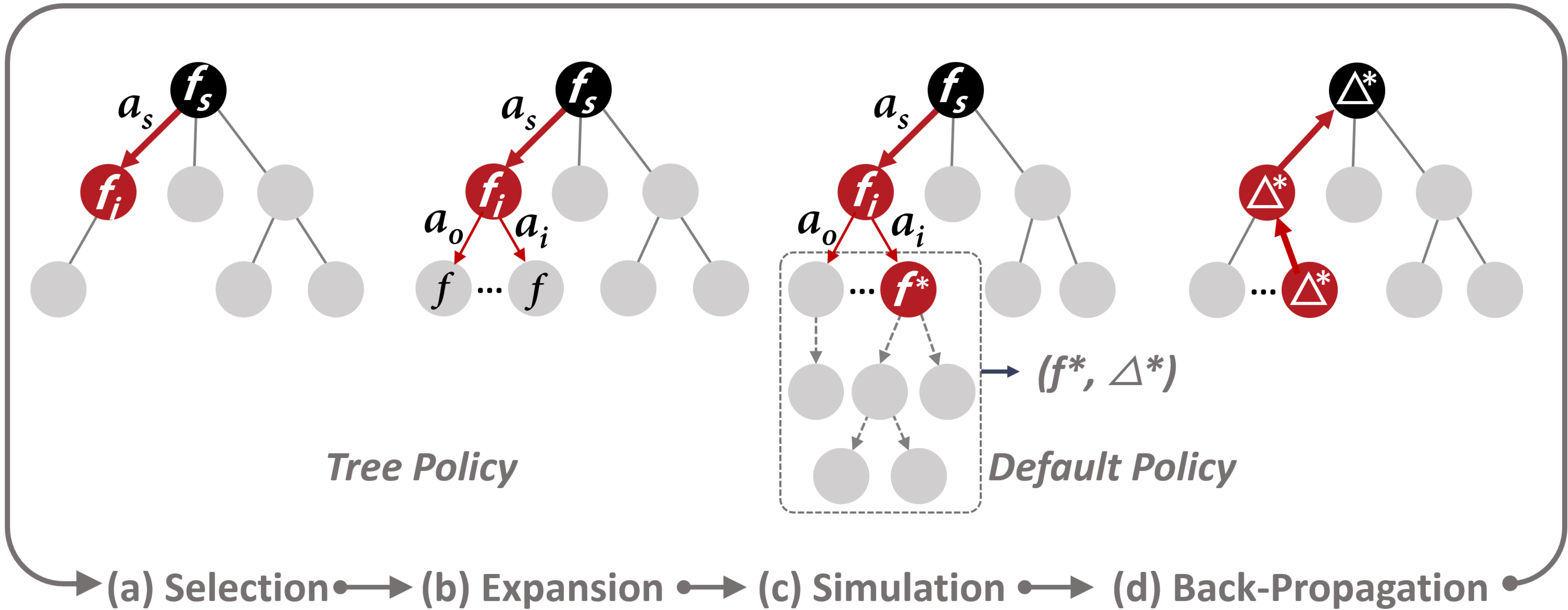}
  \vspace{-2em}
  \caption{An iteration of the Monte Carlo tree search algorithm consists of four steps, including (a) selection, (b) expansion, (c) simulation, and (d) back-propagation.} 
  \label{fig:algorithm}
  \vspace{-1.5em}
\end{figure}

\vspace{-0.25em}
\paragraph{\bf Loss Function} To yield a meaningful fact embedding ({\bf C1}) and interpolation results ({\bf C2}), the following loss has been designed and used when training the model: 
\begin{equation}
L = \sum_{(v_{i-1},v_i,v_{i+1})\in D_s}{d(v_i,m_i)^2 + \alpha \cdot d(v_i,v_j)^2}
\end{equation}
where $v$ indicates the vector representation of a data fact; $D_s$ denotes the set of training samples, which are the trigrams of facts, i.e., $(v_{i-1}, v_{i}, v_{i+1})$;  $d(\cdot)$ calculates the euclidean distance between two vectors. The first term estimates the differences between the embedded vector $v_i$ and the euclidean midpoint $m_i = (v_{i-1} + v_{i+1}) / 2$. The second term estimates the length of a trigram in $D_s$. $\alpha$ balances between these two parts. Basically, this loss function tends to reduce the distance between related facts and try to line up the vectors in a trigram.

\paragraph{\bf Training Corpus}  To train the model, we selected 100 high-quality data stories that were manually authored by our workshop participants based on different datasets using the Calliope system. All of these stories consist of 5 data facts with diverse fact types. They were designed by following either the time-oriented narrative structure~\cite{lan2021understanding} or the parallel structure~\cite{shi2021autoclips}. 300 fact trigrams were extracted from these stories as our training set. Each of them consisted of 3 succeeding data facts in the original story. 


$\mathbf{Implementation}$ We implemented the above model in PyTorch. We chose Adam optimizer and updated all the training parameters with a learning rate of 0.01. The model was trained on an Nvidia Tesla-V100 (16GB) graphic card.

\subsection{Interpolator}
\label{sec:algorithm}

In \name, the interpolator is designed to interpolate between two data facts $(f_s,f_t)$ to generate new facts in the middle that semantically connect $f_s$ and $f_t$ as the new story content. The interpolation process consists of three major steps. We first convert the facts into their vector representations $v_s$ and $v_t$ based on the fact embedding technique introduced above. After that, we directly calculate the linear interpolation between vectors as follows:
\begin{equation}
v_k = v_s + \frac{k}{N+1} \cdot (v_t - v_s)
\label{eq:interpolation}
\end{equation}
\cn{where $N$ is the total number of midpoints to be calculated. It is a user input which controls the length of the resulting story}; $v_k$ ($k\in[1,\cdots,N]$) is the $k$-th vector interpolation between vector $v_t$ and $v_s$. \cn{Here, linear interpolation is chosen due to the explainability and intuitiveness of its results that make the evaluation simple.} Finally, we search through the fact space to find a data fact whose vector representation is the most similar to $v_k$ as the final output of the $k$-th fact interpolation.

In our system, we employed the Monte Carlo Tree Search algorithm (MCTS)~\cite{browne2012survey} to ensure efficient searching through the fact space to find proper interpolation results. In particular, this algorithm dynamically constructs a searching tree $\mathcal{T}$ based on a set of predefined actions to explore the vast fact
space. As shown in Fig.~\ref{fig:algorithm}(a), each node in the tree is a data fact $f_i$ and each directed edge indicates an action through which a child node is created. 

These constraints were carefully selected based on the inspection of \menndy{our embedding space characteristics} using numerous experiments. It helps us to eliminate the data facts that might be irrelevant to the story.

Here, we carefully defined a number of constrained actions for the algorithm to choose under different conditions (Table~\ref{tab:actions}). \cn{The constraints were set by inspecting the changes of a fact's vector in the embedding space after performing different actions. These constraints guarantee to generate nodes (i.e., facts) that are meaningful and closely related to their predecessor.} Generally, the MCTS iteratively runs the following four steps to search through a vast searching space until the target is reached:

\begin{enumerate}[topsep=2pt]
\itemsep -1mm
\item[(1)] select the node $f_i$ with the largest reward score in $\mathcal{T}$ (Fig.~\ref{fig:algorithm}(a)); 
\item[(2)] expand $f_i$ by creating a number of related facts via the set of predefined actions (Fig.~\ref{fig:algorithm}(b));
\item[(3)] simulate the search based on $f_i$ and its descendants to explore a few steps further, so that the different searching directions can be estimated in advance (Fig.~\ref{fig:algorithm}(c));
\item[(4)] update the reward score on each node in $\mathcal{T}$ that is calculated during the simulation (Fig.~\ref{fig:algorithm}(d)). 

\end{enumerate}

\begin{figure}[!t]
  \centering
  \includegraphics[width=\linewidth]{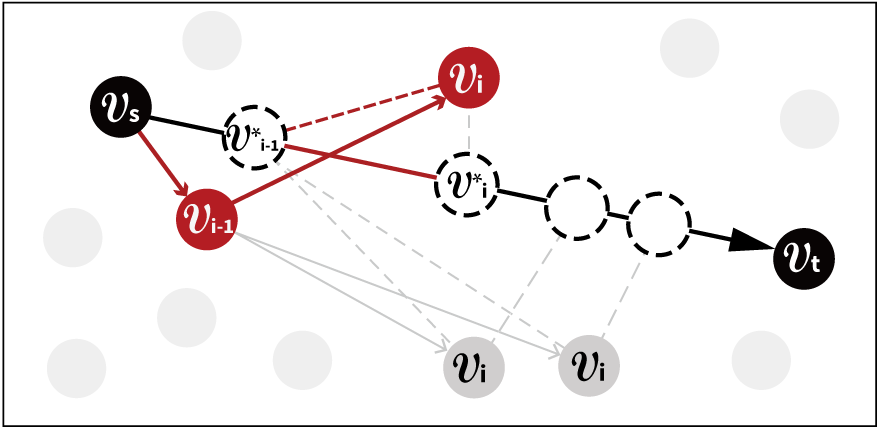}
  \vspace{-2em}
  \caption{The fact interpolation via MCTS in the vector space.} 
  \label{fig:interpolation}
  \vspace{-1.5em}
\end{figure}

A reward function is designed to estimate the quality of each searching path in the tree. The reward scores are marked on the tree nodes. They are used to guide the exploration of fact space. To ensure fast and precise searching, we define the reward to align the searching direction with the interpolation direction indicated by $\overrightarrow{v_sv_t}$ (Fig.~\ref{fig:interpolation}). Formally, the reward is defined as:
\begin{equation}
    reward(f_i) = -\frac{1}{i}\sum_{j=1}^{i}  \left | v_j-(v_{j-1}^*+\frac{v_t-v_s}{\left | v_t-v_s\right |}\cdot \left |v_j-v_{j-1}^* \right | ) \right |
\label{eq:reward}
\end{equation}
where $f_i$ is a node (i.e., a data fact) in $\mathcal{T}$ that is under estimation whose vector representation is $v_i$ as shown in Fig.~\ref{fig:interpolation}. $v_j$ in Eq.~\ref{eq:reward} is the vector representation of a node (i.e. fact $f_j$) in the searching path ending at $f_i$. $v_{j}^*$ is the expected position of $v_{j}$ on $\overrightarrow{v_sv_t}$ in the vector space. It is determined by the step length of the current search, i.e., $|v_j - v^*_{j-1}|$. Ideally, when the searching path is perfectly aligned with $\overrightarrow{v_sv_t}$, $v_{j}$ and $v_{j}^*$ will be precisely overlapped.
The above reward estimates the averaged vector distance between the actual searching path and the desired interpolation path alone $\overrightarrow{v_sv_t}$. A searching path closer to $\overrightarrow{v_sv_t}$ is encouraged. The leading negative sign is added to make the reward optimization a maximization problem.

The algorithm ends at a point when it tries to expand a node whose vector representation is close enough to the target node's vector $v_t$ in the vector space. The nodes in the searching path with the largest reward in $\mathcal{T}$ are examined. And the ones that are the closest to the midpoints calculated based on equation~\ref{eq:interpolation} are taken as the interpolation results. In this way, all the interpolated facts between $f_s$ and $f_t$ could be found through the same searching process.

\setlength{\textfloatsep}{0pt}
\begin{algorithm}[tb]
\label{alg:interpolation}
\SetAlgoLined
\SetKwInOut{Input}{Input}
\SetKwInOut{Output}{Output}
\Input{$\ {f_s}, {f_t}, {N}, {D}$}
\Output{$\mathcal{S} = \{f_s, f_1, \cdots, f_N, f_t\}$}

$\mathcal{S}\leftarrow Initialize (\{f_s, f_t\})$\;
$v_s\leftarrow embed(f_s)$; $v_t\leftarrow embed(f_t)$\; 
$v^* \leftarrow v_s$; $\mathcal{T}$ $\leftarrow$ \{$f_s$\}\;

{\color{brown}\tcp{Calculate the midpoints in the direction of the storytelling $\overrightarrow{v_s v_t}$.}}
  $I^*\leftarrow  interpolate(v_s, v_t, N)$ \;
  
\While{$\left | v^*-v_t \right |< \lambda $ }{
    {\color{brown}\tcp{ The TreePolicy consists of two steps: when the node is not fully expanded, it first selects the best child in $\mathcal{T}$ and then expands the corresponding nodes within the executable actions.}}
    $F_i\leftarrow TREEPOLICY(\mathcal{T})$\;
    
    {\color{brown}\tcp{ The DefaultPolicy is then used until the time limit has been reached. It simulates and calculates the reward to choose the node with the maximum reward.}}
    $\Delta^*,f^* \leftarrow DEFAULTPOLICY(F_i)$\;
    
    {\color{brown}\tcp{Update the reward score on each node in $\mathcal{T}$ that is calculated during the simulation.}}
    $BackPropagation(\mathcal{T},f^*, \Delta^*$)\;
    $v^*\leftarrow embed(f^*)$\; 
}
     {\color{brown}\tcp{Choose the best path  in $\mathcal{T}$ that matches the midpoints as a data story.}}
    $\mathcal{S}= \{f_s, f_1, \cdots, f_N,f_t\}\leftarrow Match(I^*,\mathcal{T})$\;
\Return $\mathcal{S}$\;
\caption{Interpolation Algorithm based on MCTS}
\end{algorithm}

\begin{figure*}[!th]
  \centering
  \includegraphics[width=\linewidth]{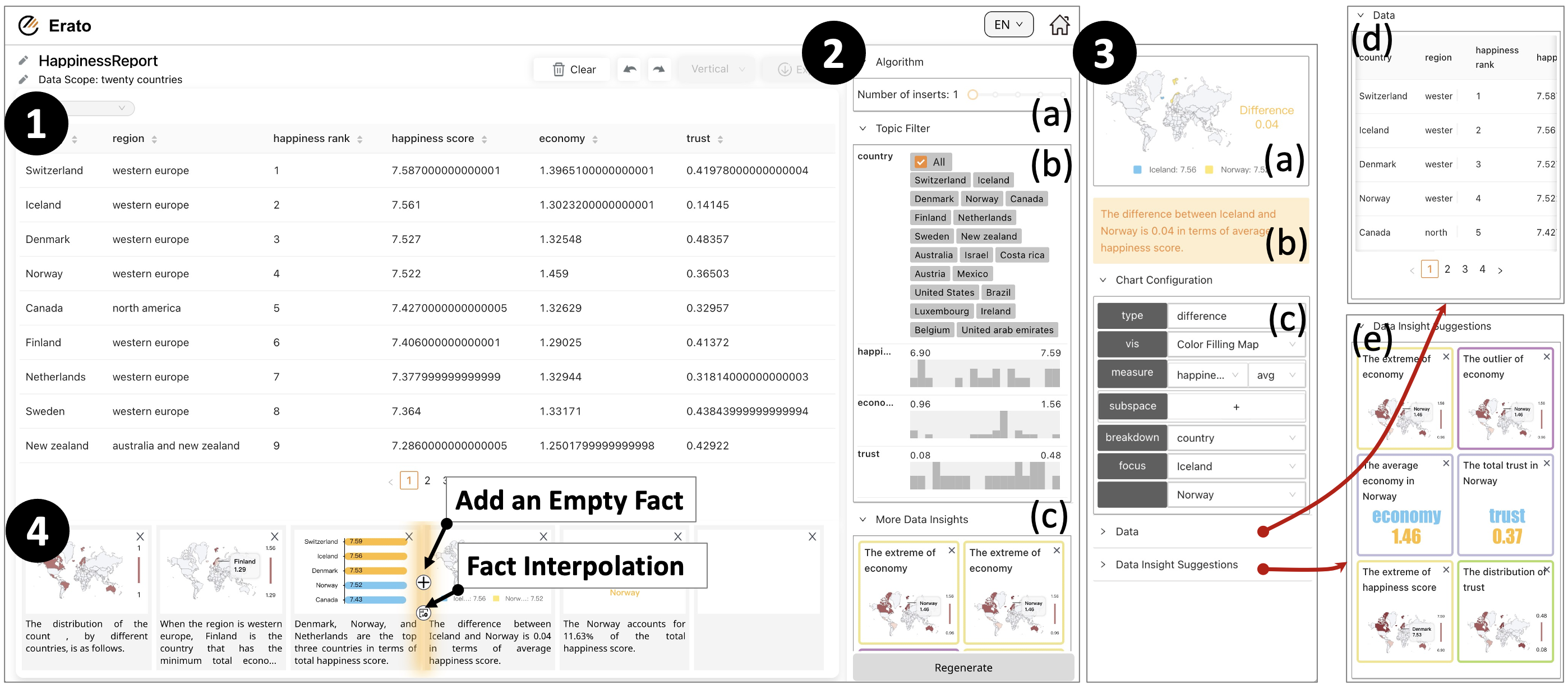}
  \vspace{-1.5em}
    \caption{The interface consists of four major components: the data story view \Circled[fill color=black, inner color=white]{1}, two configuration panels for the story mode \Circled[fill color=black, inner color=white]{2} and the fact mode \Circled[fill color=black, inner color=white]{3}, respectively, and the storyline view \Circled[fill color=black, inner color=white]{4}.}
  \label{fig:ui}
  \vspace{-1.5em}
\end{figure*}

\vspace{-0.25em}
\paragraph{\bf Algorithm Overview} Alg.~\ref{alg:interpolation} summarizes the above ideas in pseudo-codes under the MCTS's algorithm framework. In particular, the algorithm takes a pair of endpoint facts ($f_s$, $f_t$), the total number of midpoints to be interpolated $N$, and a spreadsheet $D$ as the inputs. A set of $N$ data facts that meaningfully connects $f_s$ and $f_t$ are generated as the output. Initially, $f_s$ and $f_t$ are added into an empty set $\mathcal{S}$ (\textit{line 1}) and are converted into their vector representations $v_s$ and $v_t$ based on our fact embedding technique (\textit{line 2}). We use $v^*$ to indicate the vector representation of the data fact that \smd{is} under exploration, which is initially set to $v_s$ and set the corresponding data fact $f_s$ as the root of the search tree $\mathcal{T}$ (\textit{line3}). After that, we calculate the linear interpolation between $v_s$ and $v_t$ in the vector space and store all the resulting midpoints in $I^*$ (\textit{line4}). Next, the algorithm searches through the fact space to find a set of best fits to $I^*$ via three major steps: TreePolicy($\cdot$), DefaultPolicy($\cdot$), and BackPropagation($\cdot$). In particular, \smd{t}he TreePolicy($\cdot$) selects a node with the largest reward in $\mathcal{T}$ and expends it by creating a set of data facts (denoted as $F_i$) as its children using the actions summarized in Table.~\ref{tab:actions} (\textit{line 6}, Fig.~\ref{fig:algorithm}(a,b)).
The DefaultPolicy($\cdot$) simulate\smd{s} the search based on $f \in F_i$ to explore the fact space a few steps further to find the best searching direction $f_i \rightarrow f^*$, where $f^* \in F_i$ has the largest reward $\Delta^*$ (\textit{line 7}, Fig.~\ref{fig:algorithm}(c)). The reward $\Delta^*$ of $f^*$ is then back-propagated to all the relevant nodes in $\mathcal{T}$ and $f^*$ is added into $\mathcal{T}$ as a child of $f_i$ (\textit{line 8}, Fig.~\ref{fig:algorithm}(d)). 
The above process is iteratively processed until the current best node $f^*$ is close enough to the target node $f_t$.
Finally, the searching path in $\mathcal{T}$ with the largest reward is used to make a match with the midpoints in $I^*$ (\textit{line 11}), and the set of facts that are closest to the midpoints in $I^*$ are returned in order as the interpolation results.

%% file: tables/actions.tex
\begin{table*}[!ht]
\caption{Definitions of the 7 constrained actions under different conditions, where NUM indicates the number of values or fields in the attribute. $f_s,f_t$ represent the selected pair of keyframes. The action can be performed only when the condition is satisfied.}
\label{tab:actions}
    \small
    \setlength\aboverulesep{0pt}
    \setlength\belowrulesep{0pt}
    \def\arraystretch{1.1}
    \begin{tabular}{llll}
\toprule
\textbf{Action Name}    & \textbf{Condition}   & \textbf{Description}   & \textbf{Goal} 
\\\midrule
modifyBreakdown & Breakdown($f_s$)$\ne$ Breakdown($f_t$)& Change the breakdown from $field_1$ to $field_2$ & To approach Breakdown($f_t$).                                 \\\hline

modifyMeasure & Measure($f_s$)$\ne$ Measure($f_t$)& Change the measure from $field_1$ to $field_2$ & To approach Measure($f_t$).                                         \\\hline

modifySubspace & \begin{tabular}[c]{@{}l@{}}Subspace($f_s$)$\ne$ Subspace($f_t$)\;\& \\ NUM(Subspace($f_s$))$=$NUM(Subspace($f_t$))\end{tabular} &\begin{tabular}[c]{@{}l@{}} Change the subspace from $field_1$ to $field_2$ or\\ from $value_1$ to $value_2$ \end{tabular}& To approach $f_t$'s data scope.                     \\\hline

modifyFocus & Focus($f_s$)$\ne$ Focus($f_t$)\;\& Focus($f_s$)$\ne$ Subspace($f_t$)& Change the focus from $value_1$ to $value_2$ & To approach Focus($f_t$).                                         \\\hline

\multirow{3}*{modifyType}
& \begin{tabular}[c]{@{}l@{}} $f_t$ has focus \;\& NUM(Subspace \\ ($f_s$))=0\;\& NUM(Subspace($f_t$))=0 \end{tabular}  
&Assign Focus($f_t$) to subspace  
&\multirow{2}* {\begin{tabular}[c]{@{}l@{}}To make the facts more diverse \\ and zoom the fact into \\the same data scope.\end{tabular}}\\
\cline{2-3}
~& \begin{tabular}[c]{@{}l@{}} $f_t$ has subspace \;\& NUM(Focus \\ ($f_s$))=0\;\& NUM(Focus($f_t$))=0 \end{tabular}  &   Assign Subspace($f_t$) to focus & ~ \\
\cline{2-4}
~& Except two conditions above & Change the type from $type_1$ to $type_2$ & To make the facts more diverse.

\\\hline

 addSubspace & NUM(Subspace($f_s$))$>$ NUM(Subspace($f_t$)) & Shrink the scope of Subspace($f_s$) via adding  more constraints&\multirow{2}{*}{ To approach $f_t$'s data scope.}\\
\cline{1-3}
removeSubspace & NUM(Subspace($f_s$))$<$ NUM(Subspace($f_t$)) & Enlarge the scope of Subspace($f_s$) via removing constraints &~
    \\\bottomrule    
    
\end{tabular}
\end{table*}

%% file: sections/05-interface.tex
\section{Story Editor}
In this section, we introduce the design of the story editor, which aims to provide an intuitive experience for users to explore, refine the interpolated facts, and eventually assemble them into a fluent story. \cn{The story editor, as shown in Fig.~\ref{fig:ui}, consists of four connected views}.

Once a user uploads a spreadsheet into the system, the raw data is displayed in the data story view (Fig.~\ref{fig:ui}-1). Then, the user can explore the data and select the data fields and elements that he/she is interested in by setting data filters in the configuration panel under the story mode (Fig.~\ref{fig:ui}-2(b)). Based on the selected data corpus, the underlying system analyzes and recommends a set of important data facts to help users quickly understand the data and inspire them to create meaningful and interesting stories  (Fig.~\ref{fig:ui}-2(c)), which avoids the cold start problem. 

To create a data story, a user first needs to specify a sequence of key facts and arrange them in the storyline view (Fig.~\ref{fig:ui}-4). In particular, the user can either manually insert an empty fact anywhere in the storyline \cn{(e.g., the empty fact shown in Fig.~\ref{fig:ui}-4)} or simply instruct the system to interpolate between any two succeeding facts in the storyline. When a data fact $f_i$ is selected from the storyline, the fact configuration panel is displayed (Fig.~\ref{fig:ui}-3), in which the user can edit the fact's text descriptions (Fig.~\ref{fig:ui}-3(b)) and its \type, \measure, \breakdown, \subspace, and \focus fields (Fig.~\ref{fig:ui}-3(c)). A preview of the fact is shown on top of the panel as a visualization chart (Fig.~\ref{fig:ui}-3(a)) and the data snippet corresponding to the fact is also shown in a table at the bottom (Fig.~\ref{fig:ui}-3(d)). The preview and table provide details and guide the user to create a proper data fact for the story. When the selected fact $f_i$ has a predecessor $f_{i-1}$ and a successor $f_{i+1}$ in the storyline, our system will automatically interpolate between $f_{i-1}$ and $f_{i+1}$. All valid facts on the interpolation path are displayed in a recommendation list (Fig.~\ref{fig:ui}-3(e)) as they are the candidates that could be potentially used to replace $f_i$, thus providing users with more story ideas.

Finally, the generated data story is shown in the data story view (Fig.~\ref{fig:ui}-1) in a form of a storyline (Fig.~\ref{fig:teaser}), a factsheet (Fig.~\ref{fig:factsheet}), or a scroll-up view (Fig.~\ref{fig:swiper}). The user can easily switch between different representation forms via a drop-down menu. \cn{The story view is implemented based on the library released by the Calliope project~\cite{calliope:chart}. In particular, a chart library is used for visualizing individual facts and a data story library is used for \smd{showing} stories in the \smd{aforementioned} different forms.}





%% file: sections/06-evaluation.tex
\section{Evaluation}
We evaluated \name and the corresponding key techniques via both quantitative experiments and controlled user studies. In particular, we first verified the consistency between the fact's vector representation and human cognition via a user study. Next, we evaluated the interpolation results via a quantitative experiment that measured their overall performance and a Turing test that estimated the quality of the generated content from the human perspective.
Finally, three case studies together with domain expert interviews were conducted to verify the overall usability of the system. 

\subsection{Evaluation of the Fact Embedding Results}
We conducted a user study to verify the consistency of the fact similarities calculated based on the facts' embedding vectors and the fact similarities perceived by users \smd{regarding} to the facts' semantics. A consistent result indicated that our embedding technique was able to successfully capture the fact semantics and their relationships.

\underline{\textit{Procedure and Tasks.}}
We first prepared 30 manually generated data facts for the study. Each fact, $f_i$, was accompanied by two other randomly generated data facts ($f_a$, $f_b$) whose cosine similarities to $f_i$ were calculated as the ground truth. \cn{We ensured the difference between $f_a$ and $f_b$ was smaller than 0.25 to check if the participants were sensitive to small differences.} As a result, 30 triplets of data facts were prepared. In the study, we showed one triplet at a time to a participant and asked him/her to identify the fact closer to $f_i$ based on his/her own judgments. Their answers were recorded and the corresponding accuracy \cn{(i.e., the percentage of right answers)} was calculated. 

\cn{To make sure the participants fully understand the data insight captured by a fact, each fact was shown in a visualization chart with a caption manually written by us in an offline procedure}. To ensure a fair and comprehensive comparison, these facts were generated based on \smd{6 datasets covering 6 different topics, including public health, politics, economy, sports, recreation, and industry which} had similar schema and data distributions. We also counterbalanced the fact types. During the test, we encouraged the participants to first understand the data insights before providing their answers. 30 participants (7 males and 23 females, mean age 24) were involved in this study. It took an average of 15 minutes for a participant to finish all 30 triplets.

\underline{\textit{Results.}} 
On average, the accuracy was 89\% with a standard deviation of 0.08. This result suggested that in most testing cases, the participants agreed that the computationally more similar facts were also perceivably more similar. It verified that our embedding algorithm was consistent with human cognition.

\begin{figure}[t]
  \centering
  \vspace{-0.5em}
  \includegraphics[width=0.9\linewidth]{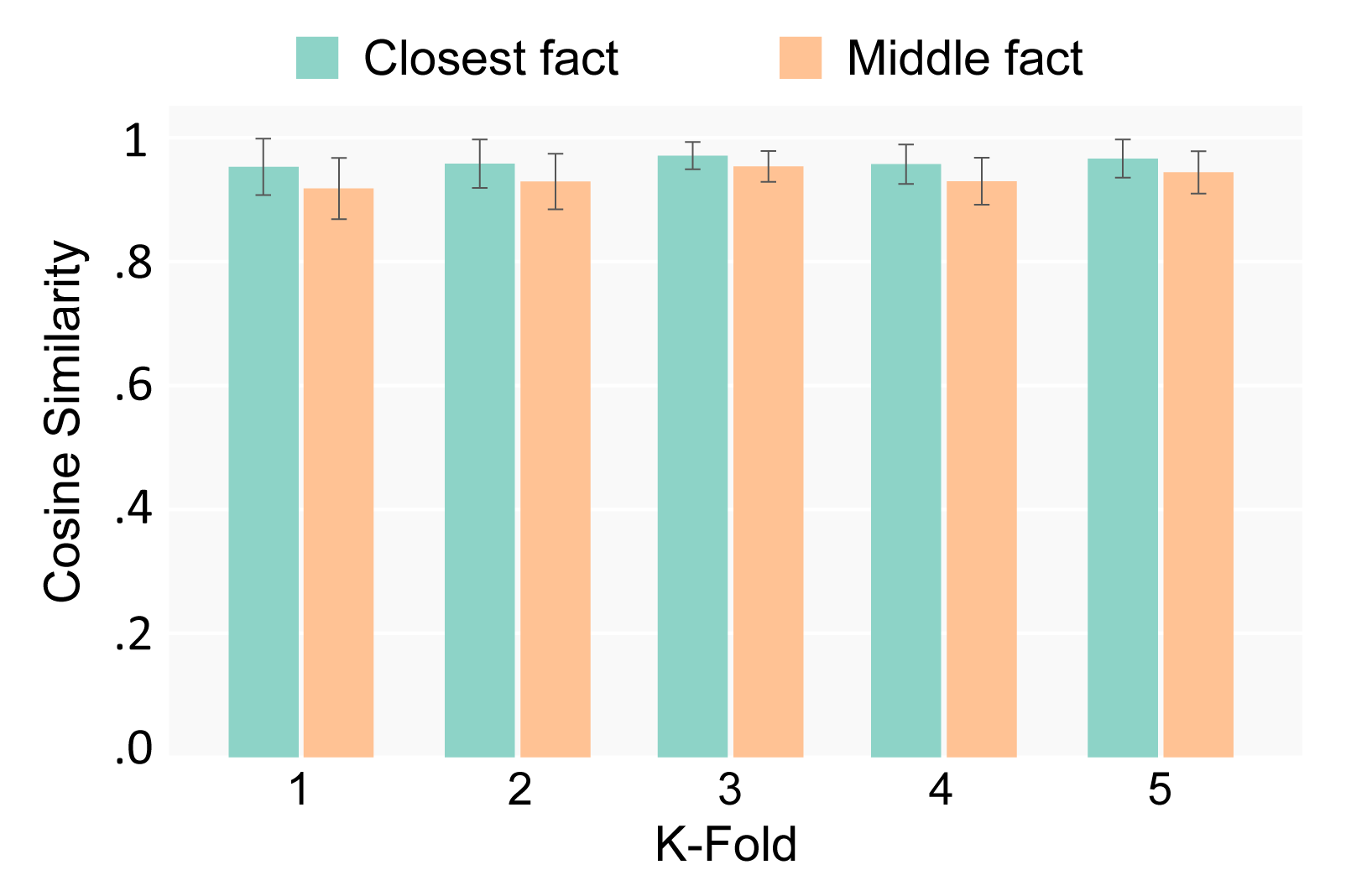}
  \vspace{-1em}
  \caption{ The 5-fold cross-validation of the interpolation technique.} 
  \label{fig:5-fold_validation}
  \vspace{-1.3em}
\end{figure}
\setlength{\textfloatsep}{15pt}
\subsection{Evaluation of the Fact Interpolation Results}
We estimated the interpolation technique through a quantitative evaluation to check the precision of the interpolation results. A Turing test was also performed to verify that the results also aligned with a human's perception.

\paragraph{\textbf{\textit{Quantitative Evaluation}}}
We performed 5-fold cross-validation to estimate the performance of the proposed interpolation technique based on the dataset introduced in Section~\ref{sec:model}. In particular, the dataset was first divided into five equal folds, each containing 60 fact trigrams. In the next, we used four folds of data to train the fact embedding model but left the rest one fold for testing the performance of the interpolation technique based on the embedding model. The whole training and testing process was interactively performed five times with a fold of training and testing data shifting at each time.

The training phase was to fine-tune BERT based on our loss function as described in Section~\ref{sec:model}. During the testing phase, we interpolated between the first fact and the last fact in a trigram in the testing set. The interpolation results, i.e., a sequence of data facts $S = \{f_1,\cdots,f_n\}$, were compared to the ground truth (the second fact in the trigram) in terms of cosine similarity via two strategies: (1) comparing with the middle fact $f_{\lceil n/2 \rceil}$; (2) comparing with the closest fact, i.e., the fact in $S$ that was the most similar to the ground truth. On average, the similarities for strategy (1) and (2) were 0.944 ($\sigma=0.034$) and 0.953 ($\sigma=0.045$), respectively. The cross-validation results were shown in Fig.~\ref{fig:5-fold_validation}. 

\paragraph{\textbf{\textit{Turing Test}}}

To verify that our algorithm was able to create high-quality results from a user's perspective, we also performed a Turing test to let users differentiate the data stories that were fully generated by human designers (denoted as $G_h$) from those that were partially generated by our interpolation algorithm (denoted as $G_m$). We established the following hypotheses:
\begin{enumerate}[topsep=4pt,parsep=4pt]
\itemsep -1mm
\item[{$H1$}] There was no significant semantic difference between the human-generated ($G_h$) and interpolation-based ($G_m$) results.
\item[{$H2$}] $G_m$'s quality was as good as that of $G_h$ in terms of the coherence of the corresponding data stories.
\end{enumerate}

\underline{\textit{Procedure and Tasks.}}
The Turing test consisted of two stages. In the first stage, we invited 2 senior graduate students to manually create short data stories consisting of 5 data facts using our system with the interpolation feature disabled. Both students were female from a top design college and had rich \smd{experience} in designing visual data stories. 
\smd{They were provided with 15 datasets covering 5 different topics, including public health, society, economy, sports, and recreation}. In total, 15 data stories ($G_h$) were created based on these datasets with diverse and balanced topics (3 stories for each \smd{topic}). After that, we replaced the middle three data facts of each data story with another three data facts generated by interpolating between the first and the last data facts based on the proposed technique. As a result, another 15 stories ($G_m$) were created. Here, we controlled the story length as five to make it not too short so that the interesting content could be captured but also not too long so that the participants could easily read them without spending too much time. We chose to replace three middle facts as we would like to replace the human-generated facts in each data story as many as possible, so that we could clearly check whether the facts generated by our algorithm would affect the coherence of the data story or not. 

\begin{figure}[tb]
  \centering
  \vspace{-0.5em}
  \includegraphics[width=0.85\linewidth]{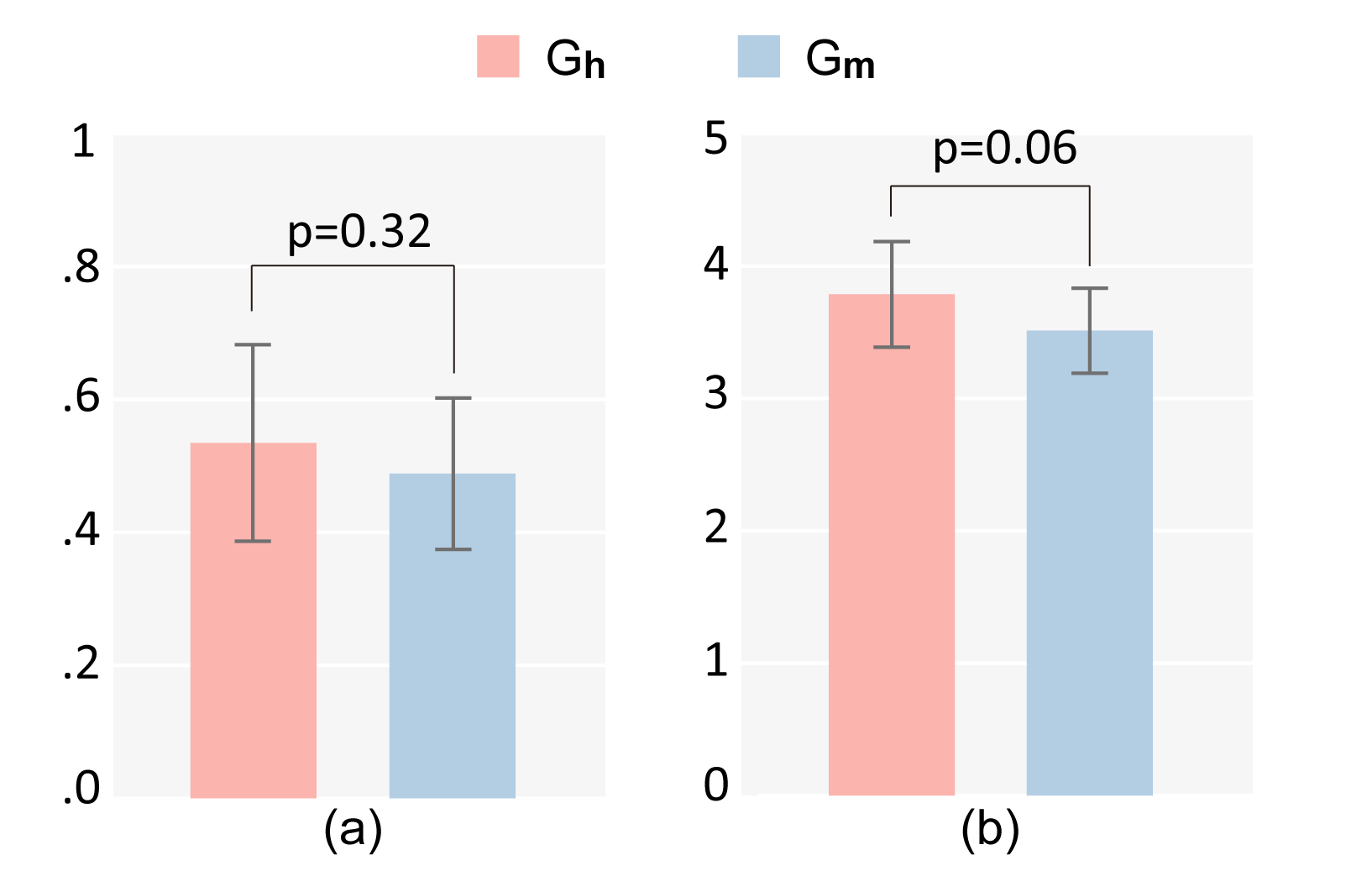}
  \vspace{-1em}
  \caption{Means and standard errors of each item. (a) Percentage of stories perceives as human, (b) ratings on the quality of data stories using a 5-point Likert scale.} 
  \label{fig:Turing}
  \vspace{-1.3em}
\end{figure}

In the second stage. we randomly mixed all 30 stories together and put them into an online questionnaire. Another 50 participants (18 males and 32 females, mean age 28.78) were invited to differentiate these data stories. These participants had diverse backgrounds. Some of them were university students majoring in design, architecture, computer science, and mathematics. Some of them were employees in an IT consulting company, faculty members in a university, and data journalists from a news media. In the questionnaire, we showed one story at a time to a participant, who was asked to finish the following two tasks designed regrading to the above hypotheses:
\begin{enumerate}[topsep=4pt,parsep=4pt]
\itemsep -1mm
\item[{$T1$}] Tell whether the story was fully generated by a human or partially generated by a machine regarding its content.
\item[{$T2$}] Rate the quality of the story in terms of its logic coherence using a 5-point Likert scale with 5 indicating the best quality.
\end{enumerate}
To facilitate understanding, each fact was presented as a visualization accompanied by a manually-written caption.
At the end of each test, we encouraged the participants to leave the reasons for their decisions, so that we could find the potential limitations of our technique. On average, each participant spent about 20 minutes completing the test.

\underline{\textit{Results.}}
We first reported the accuracy of each group and then discussed the feedback from the participants.

\textit{Accuracy.} Fig.~\ref{fig:Turing}(a) showed the results of the first task ($T1$), where y-axis indicated the percentage of positive ratings (i.e., identified as fully human generated). Not surprisingly, the performance of $G_h$ ({\it M} = 0.53, {\it SD} = 0.16) was better than that of $G_m$ ({\it M} = 0.48, {\it SD} = 0.13). However, a paired t-test showed that the difference was not significant ($\alpha = 0.05$, $p = 0.32$), thus $H1$ accepted. Fig.~\ref{fig:Turing}(b) showed the results of the second task ($T2$), where y-axis indicated the Likert rating on the data story quality. Again, there was no significant difference ($\alpha = 0.05$, $p = 0.06$) between two groups, but $G_h$ had a better average rating \smd{{\it Human}} ({\it M} = 3.79, {\it SD} = 0.39) than that of $G_m$ \smd{{\it \name}} ({\it M} = 3.51, {\it SD} = 0.32)(Fig.~\ref{fig:Turing}(b)), thus $H2$ accepted as well.

\textit{Feedback.}
During the study, almost all the participants indicated that the data stories were very subjective and they relied primarily on their intuitions to make inferences. They also left their reasons and comments on their choices which were summarized as follows: 

\begin{itemize}[leftmargin= 10pt,topsep=1pt,parsep=4pt]
\itemsep -.5mm
\item \textit{Logicality}, regarding the narrative structures. Three participants pointed out that stories created by people tended to have more complex narrative structures, such as the three-act structures, but ``the algorithm tends to generate a parallel structure". At the same time they also mentioned ``it is difficult to differentiate stories with a time-oriented structure". We believed this was because it was easier for the algorithm to interpolate on the temporal dimension to generate similar and parallel content. At the same time, we acknowledged that taking narrative structure into consideration was indeed a missing part of the proposed algorithm.

\item \textit{Diversity}, regarding the variety of the fact types and the complexity of \smd{the} data content. Several participants believed stories generated by humans should contain rich fact types which made the whole story vivid. Therefore, they tended to judge stories that contained duplicate content or data facts of the same type as machine-generated. This finding reminded us that even for an interpolation task, the diversity of the content was as important as logical smoothness.

\item \textit{Meaningfulness}, regarding the meaning of telling a story.  A number of participants mentioned that the story of human creation might connote a certain trend and \smd{allow} readers to draw some conclusions from it. One participant pointed out that ``it will be more likely to be generated by a human if the story is thought-provoking". In addition, some participants believed ``[people] not tend to illustrate data facts of common \smd{senses}". We believed generating insightful stories was a great challenge for a fully automated algorithm, which showed the value of human-machine collaboration. 

\end{itemize}

\subsection{Interview with Experts}
To further evaluate the usability of \name, we conducted a semi-structured interview with three domain experts (denotes by \textbf{E1}-\textbf{E3}). 
The first expert was a data analyst \smd{with 3 years of working experience, whose major job was to analyze customer data}. The second expert, a senior designer, \smd{had 5 years of experience in creating infographics}. The third expert was a data journalist who had more than 4 years of working experience and was familiar with data storytelling and data story authoring. 

\vspace{-0.4em}
\paragraph{\textbf{\textit{Datasets}}}
We collected three datasets covering three different topics:
natural environment (\textbf{D1}), entertainment \smd{(\textbf{D2})}, and sports competition (\textbf{D3}). Specifically, \textbf{D1} contained all the natural disasters worldwide since 2000 (8958 rows, 9 columns). It recorded disaster types, subtypes, year, month, country, region, continent, the number of deaths, and affected people. \textbf{D2} contained all Disney films produced since 1937 (375 rows, 9 columns), including the movie's title, genre, year of release, country, language, running time, box office, and IMDB ratings. \textbf{D3} recorded the number of gold, silver, \smd{and} bronze medals won by a country in each type of sport during the 2022 Winter Olympics (118 rows, 6 columns). 
These three datasets were used for the case study and had been respectively distributed to the experts.
\begin{figure}[t]
  \centering
  \vspace{-0.5em}
  \includegraphics[width=\linewidth]{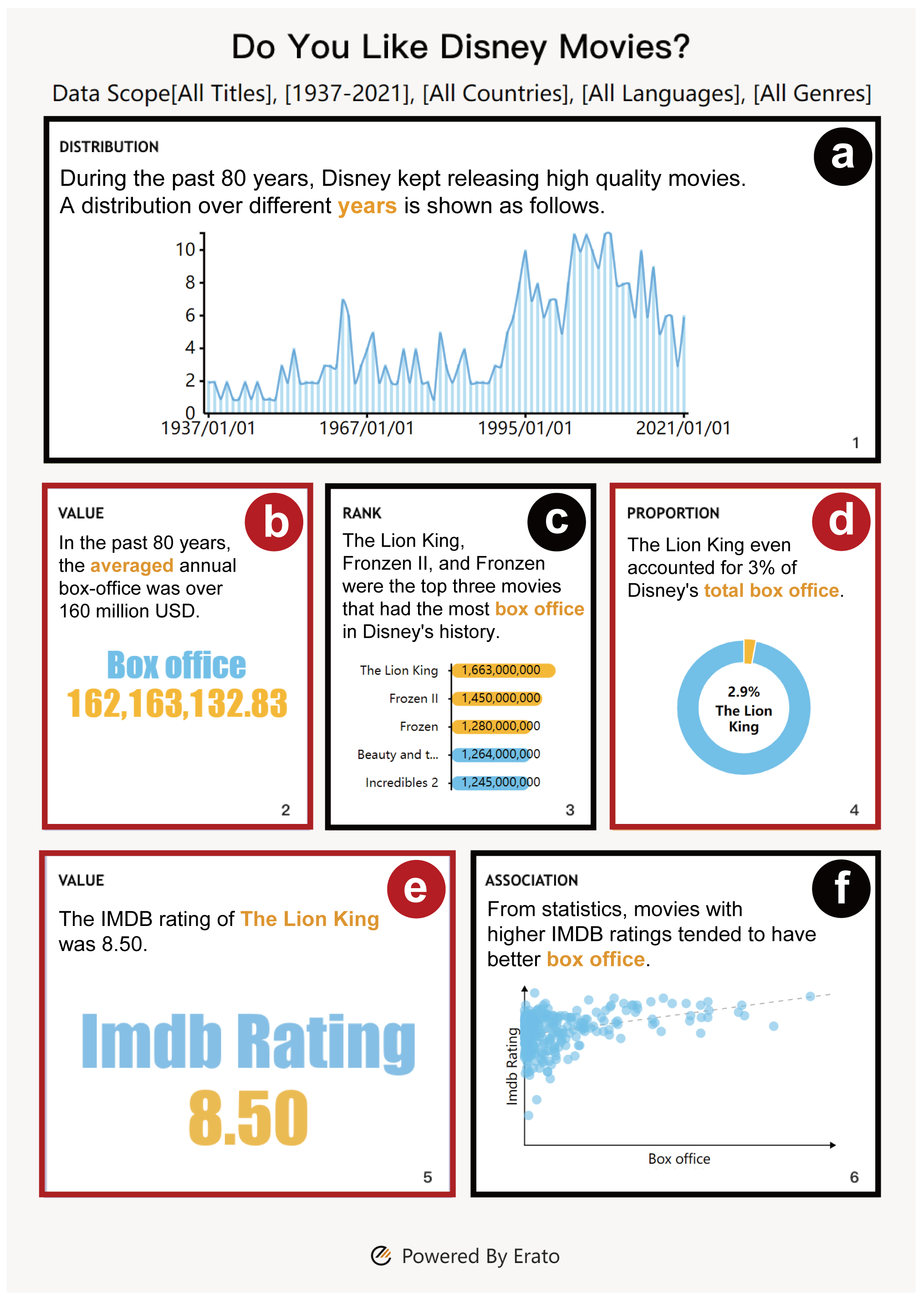}
  \vspace{-2em}
  \caption{A data story about Disney movies represented in form of a \factsheet that was authored by an expert user using \name during our interview. The data facts (a, c, f) were created by \menndy{the user, based} on which the facts (b, d, e) were generated based on the proposed interpolation technique. The story first shows the number of films released each year over the past 80 years (a), followed by the average annual box office (b) and the corresponding ranking of films (c). It gradually focuses on the most popular movie "The Lion King" (d, e), and concludes that the IMDB rating is positively correlated with the box office (f).} 
  \label{fig:factsheet}
  \vspace{-1.3em}
\end{figure}
\begin{figure}[ht]
  \centering
  \vspace{-0.5em}
  \includegraphics[width=\linewidth]{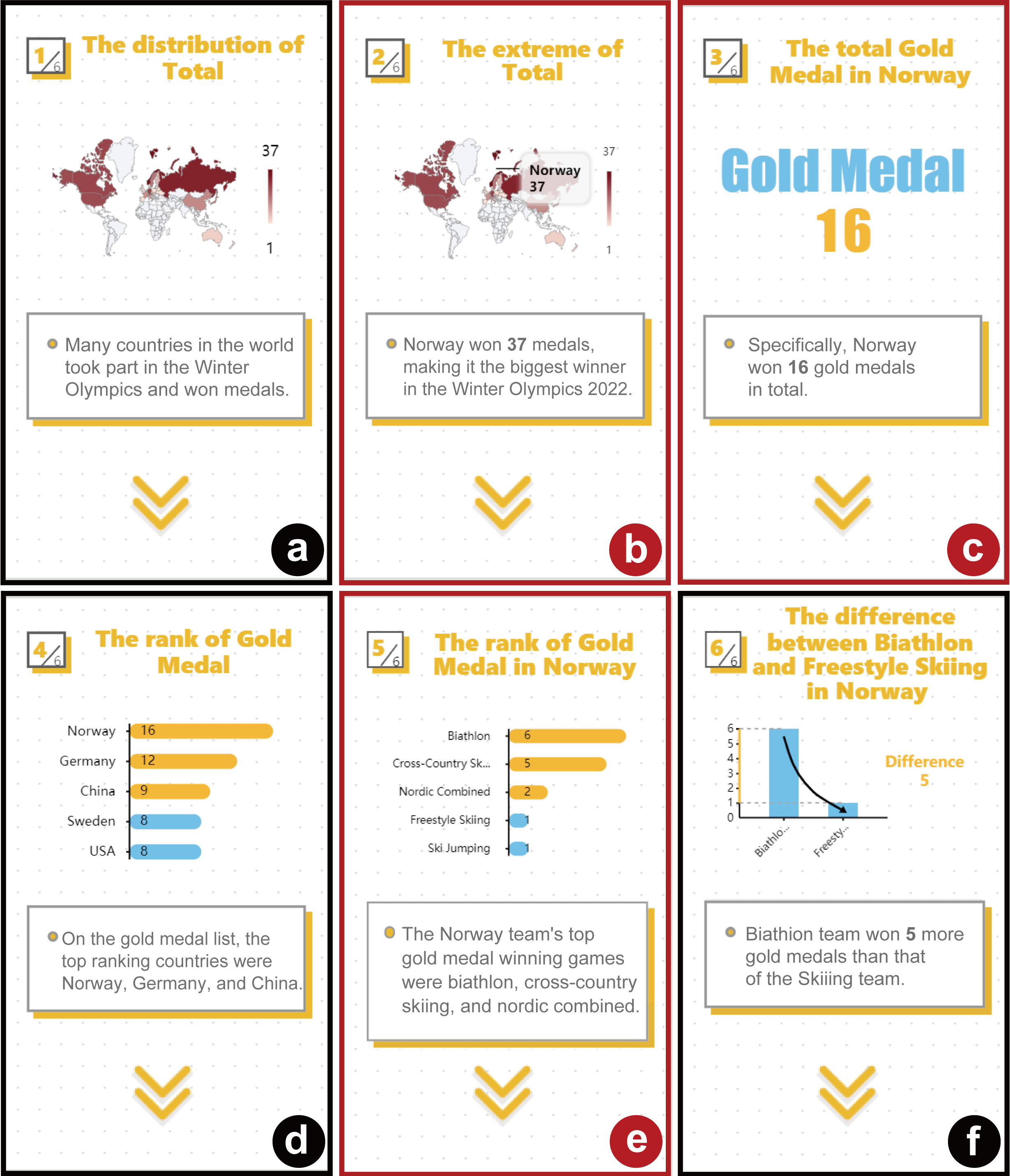}
  \vspace{-1em}
  \caption{A data story about the Winter Olympic Games 2022 represented in form of an scroll-up view, which is created by a data journalist using \name during our interview. The data facts (a, d, f) were created by the journalist as the keyframes for the story, based on which  (b, c, e) were generated based on our interpolation technique. The story first illustrates an overall geographical distribution of medal-winning countries (a) followed by an elaboration of specific data from Norway, the top-ranked country in the Winter Olympics (b-e). The story finally makes a comparison between biathlon and freestyle skiing and reveals Norway's strength in biathlon (f).} 
  \label{fig:swiper}
  \vspace{-1.3em}
\end{figure}

\vspace{-0.4em}
\paragraph{\textbf{\textit{Procedure and Tasks}}}
Because of the COVID-19 pandemic, all the interviews were conducted online. At the beginning of each interview, we introduced the purpose of our study as well as the data content. We briefly demonstrated \name system. Each expert then spent 10 minutes familiarizing him/herself with \name and was asked to use it to create a data story consisting of six data facts with the given dataset based on \name. All the experts were encouraged to think aloud during the creation process. In order not to interfere with their creative thinking, we did not set any time limit to the process. The experts might work as long as they want. We saved the final stories created by these experts. \smd{On average, it took about 30 minutes for an expert to create a story.} After creating the stories, interviews were performed separately to collect their comments on three aspects: (1) the coherence of the interpolated facts and the usefulness of the technique; (2) the overall quality of the generated data stories, and (3) the usability of \name. Each interview lasted for about one hour with the processes recorded for later analysis.

\vspace{-0.4em}
\paragraph{\textbf{\textit{Study Results}}}
We first reviewed the stories generated by our expert users during the case study and then presented their comments on our techniques and systems that were collected during the interview.

\underline{\textit{The authored data stories.}} Three data stories authored by our experts were illustrated in Fig.~\ref{fig:teaser}, Fig.~\ref{fig:factsheet}, and Fig.~\ref{fig:swiper}. In these data stories, the data facts generated by interpolation were marked in red, whereas the keyframes input by users were marked in black. 

Fig.~\ref{fig:teaser} shows a data story entitled ``Nature is Sounding the Alarm" that was created by \textbf{E1} based on the natural disaster dataset (\textbf{D1}). It consists of four keyframes (\textit{Fact a, c, e, f}) and two interpolated data facts (\textit{Fact b, d}). Specifically, over the past decades, a number of regions suffered from natural disasters (\textit{Fact a}). Geographic distribution of their occurrences showed how frequently these regions had been affected (\textit{Fact b}). The top three most influenced areas were Asia, America, and Africa (\textit{Fact c}). Nearly 40\% of disasters occurred in Asia (\textit{Fact d}), with China having the highest incidence of disasters (\textit{Fact b}). The three most frequently occurring natural disasters in China were tropical cyclones, river floods, and earthquakes (\textit{Fact e}), with flooding having the most serious impact (\textit{Fact f}), which needed more attention.

Fig.~\ref{fig:factsheet} presents a story about Disney movies (\textbf{D2}) created by \textbf{E2}, which consists of three keyframes (\textit{Fact a, c, f}) and three interpolated facts (\textit{Fact b, d, e}).
In particular, during the past 80 years, Walt Disney has released a number of films. Especially after 1995, the number of films released a year increased dramatically but gradually decreased in recent years (\textit{Fact a}). On average, the annual box office was over 160 million USD (\textit{Fact b}). The top three films with the best box office were the Lion King, Frozen II, and Frozen (\textit{Fact c}), and the best one, the Lion King even took about 3\% of total box offices in Disney's history (\textit{Fact d}). Its IMDB rating was 8.5, which was quite a high score (\textit{Fact e}). The statistics showed that the IMDB rating was positively correlated with the box offices (\textit{Fact f}).

Fig.~\ref{fig:swiper} illustrates a data story about the Winter Olympic Games 2022 held in Beijing (\textbf{D3}), which was created by \textbf{E3}. The story included three keyframes (\textit{Fact a, d, f}) and three interpolated facts (\textit{Fact b, c, e}). Specifically, many countries in the world took part in the game and won a medal (\textit{Fact a}). Among all these countries, Norway won the most medals, making it an outlier (\textit{Fact b}). The total number of gold medals in Norway was 16 (\textit{Fact c}). It ranked number one on the Olympic gold medal list (\textit{Fact d}). In addition, the three strongest sports of the Norwegian team were biathlon, cross-country skiing, and nordic combined. Norway won the most gold medals in biathlon (\textit{Fact e}), with five more medals than in freestyle skiing (\textit{Fact f}) which also revealed Norway's strength in biathlon.

\underline{\textit{Interview Feedback.}}
In the follow-up interview, the expert provided a number of valuable comments that were summarized as follows:

\textit{The usefulness of the interpolation.}
All the experts agreed the proposed interpolation technique was very helpful in terms of supporting both data exploration in context (\textbf{E1}) and data story editing (\textbf{E2,3}). In particular, \textbf{E1} mentioned ``this feature [interpolation] is able to provide meaningful insights in the context ... is better than the quick insight features provided in other BI tools that can only generate random insights". \textbf{E2} was also impressed by our interpolation algorithm. She felt ``it is a smart function that helps complete a smooth data story". She also mentioned ``it [the interpolation technique] facilitates the ideation process, ..., especially when I haven't figured out how to tell a story." \textbf{E3} mentioned ``it [the interpolation feature] indeed saved many of my data exploration efforts when creating a story". 

\textit{The quality of the interpolation results.}
All the experts were satisfied with and impressed by the interpolation results.
 \textbf{E1} mentioned ``the generated logical order is reasonable and makes the data easier to understand".
 \textbf{E2} was impressed by the insightful data facts automatically generated by our interpolation algorithm. She said ``it is surprising that the system is able to suggest such meaningful content [i.e., data facts] based on my inputs".
 \textbf{E3} felt ``the resulting data facts are coherent in both logic and content". She said ``before using this tool, I didn't realize the intelligent techniques could be so useful ..., it even can create such a good story content [in an automatic process]".

\textit{The authoring tool.}
All the experts liked the idea of letting users cooperatively design a story with the help of an intelligent system based on the interpolation technique. They believed \name was an ``effective data story authoring tool, especially for users who lack experience" (\textbf{E2,3}). In particular, \textbf{E2} believed the system's interpolation and recommendation features were nice functions that ``provide necessary and helpful inspirations for authoring a data story". She also felt the design of the system was ``intuitive" and ``easy to get started quickly". \textbf{E3} said, ``I can easily create a data story with the help of the system". She mentioned ``it is very important to let users input the keyframes, which gives them a right to control [the content and structure of] the story". At the same time, she also agreed that ``interpolation feature will save users' efforts" and ``let them focus more on the important part". \textbf{E1} felt the tool was able to ``help people explore the insights in the data and support users to express their ideas".

%% file: sections/07-conclusion.tex
\vspace{-0.5em}
\section{Limitations and Future Work}
While the evaluation results indicate \name is promising to help users create insightful and fluent data stories, the system still has several limitations that were found during the implementation or mentioned by the participants during interviews. We hope to guide potential future research directions by pointing out these limitations.

\underline{\textit{Enriching Visualization.}}
\cn{\textbf{E1,2} would like to have additional formats such as slides and dashboards, to support more application scenarios. \textbf{E2,3} also felt the provided charts were rather conventional and notably simple. They would like to have more advanced visual representations to make the data stories vivid and engaging.}

\underline{\textit{Boosting the Creativity.}}
\cn{The story editor is not designed to boost users' creativity. For example, users cannot change the size and position of a chart in a \factsheet. They also cannot add icons or background images to enhance the narrative of the data story.}

\underline{\textit{Improving Performance and Quality.}}
The design and implementation of the current system have some performance bottlenecks. \cn{It usually takes about \smd{10-40} seconds \smd{to run} the interpolation algorithm, which, \smd{sometimes results} in considerable waiting time. The quality of the embedding model could be further improved by training it based on data stories with more sophisticated designs. In addition, although the proposed algorithm is able to generate meaningful data facts, it cannot create soulful stories that are able to affect readers.}

\underline{\textit{Conducting Thorough Evaluations.}} \cn{In this work, we did not compare the data stories generated based on \name with those automatically generated ones as we cannot control the story topic even using \smd{the} state-of-the-art automatic story generation technique. Second, the Turing test also has limitations as the quality of the \smd{human-generated} stories could be \smd{affected} by \name's editing functionality. More in-depth evaluations may help us identify more pain points and future directions.}

\section{Conclusion}
\cn{In this paper, we have presented \name, the first intelligent system designed for supporting human-machine cooperative data story design. The system employs a fact interpolation algorithm to create intermediate facts that smooth the transition between two succeeding data facts. The proposed technique was evaluated via a series of evaluations including a Turing test, a controlled user study, a performance validation, and interviews with expert users. The evaluation showed the proposed technique is sound and well accepted by our users. Just like the interpolation technique greatly accelerates the creation of animations, we believe the future development of the data content interpolation technique first introduced in this paper will greatly accelerate the traditional data story authoring process and the proposed fact embedding model will be extended and used in many visual content generation tasks.}

%% file: main.bbl
\begin{thebibliography}{10}

\bibitem{calliope:chart}
Calliope visual story chart library.
\newblock \url{https://www.npmjs.com/package/calliope-chart}.
\newblock Accessed July 28, 2022.

\bibitem{amini_authoring_2017}
F.~Amini, N.~H. Riche, B.~Lee, A.~Monroy-Hernandez, and P.~Irani.
\newblock Authoring data-driven videos with dataclips.
\newblock {\em IEEE Transactions on Visualization and Computer Graphics},
  23(1):501--510, 2017.

\bibitem{borkin2015beyond}
M.~A. Borkin, Z.~Bylinskii, N.~W. Kim, C.~M. Bainbridge, C.~S. Yeh, D.~Borkin,
  H.~Pfister, and A.~Oliva.
\newblock Beyond memorability: Visualization recognition and recall.
\newblock {\em IEEE Transactions on Visualization and Computer Graphics},
  22(1):519--528, 2015.

\bibitem{brehmer_timeline_2019}
M.~Brehmer, B.~Lee, N.~H. Riche, D.~Tittsworth, K.~Lytvynets, D.~Edge, and
  C.~White.
\newblock Timeline storyteller: The design \& deployment of an interactive
  authoring tool for expressive timeline narratives.
\newblock In {\em Proceedings of the Computation+ Journalism Symposium}, pp.
  1--5, 2019.

\bibitem{browne2012survey}
C.~B. Browne, E.~Powley, D.~Whitehouse, S.~M. Lucas, P.~I. Cowling,
  P.~Rohlfshagen, S.~Tavener, D.~Perez, S.~Samothrakis, and S.~Colton.
\newblock A survey of {M}onte {C}arlo tree search methods.
\newblock {\em IEEE Transactions on Computational Intelligence and AI in
  Games}, 4(1):1--43, 2012.

\bibitem{bryan2016temporal}
C.~Bryan, K.-L. Ma, and J.~Woodring.
\newblock Temporal summary images: An approach to narrative visualization via
  interactive annotation generation and placement.
\newblock {\em IEEE Transactions on Visualization and Computer Graphics},
  23(1):511--520, 2016.

\bibitem{chen2021vizlinter}
Q.~Chen, F.~Sun, X.~Xu, Z.~Chen, J.~Wang, and N.~Cao.
\newblock Vizlinter: A linter and fixer framework for data visualization.
\newblock {\em IEEE Transactions on Visualization and Computer Graphics},
  28(1):206--216, 2021.

\bibitem{chen2018supporting}
S.~Chen, J.~Li, G.~Andrienko, N.~Andrienko, Y.~Wang, P.~H. Nguyen, and
  C.~Turkay.
\newblock Supporting story synthesis: Bridging the gap between visual analytics
  and storytelling.
\newblock {\em IEEE Transactions on Visualization and Computer Graphics},
  26(7):2499--2516, 2018.

\bibitem{cui2019text}
W.~Cui, X.~Zhang, Y.~Wang, H.~Huang, B.~Chen, L.~Fang, H.~Zhang, J.-G. Lou, and
  D.~Zhang.
\newblock Text-to-viz: Automatic generation of infographics from
  proportion-related natural language statements.
\newblock {\em IEEE Transactions on Visualization and Computer Graphics},
  26(1):906--916, 2019.

\bibitem{demiralp2017foresight}
{\c{C}}.~Demiralp, P.~J. Haas, S.~Parthasarathy, and T.~Pedapati.
\newblock Foresight: Recommending visual insights.
\newblock {\em arXiv preprint arXiv:1707.03877}, 2017.

\bibitem{devlin2018bert}
J.~Devlin, M.-W. Chang, K.~Lee, and K.~Toutanova.
\newblock Bert: Pre-training of deep bidirectional transformers for language
  understanding.
\newblock {\em arXiv preprint arXiv:1810.04805}, 2018.

\bibitem{dibia_data2vis_2019}
V.~Dibia and C.~Demiralp.
\newblock Data2vis: Automatic generation of data visualizations using
  sequence-to-sequence recurrent neural networks.
\newblock {\em IEEE Computer Graphics and Applications}, 39(5):33--46, 2019.

\bibitem{dove2012narrative}
G.~Dove and S.~Jones.
\newblock Narrative visualization: Sharing insights into complex data.
\newblock {\em Interfaces and Human Computer Interaction (IHCI)}, pp. 21--23,
  2012.

\bibitem{fulda_timelinecurator_2016}
J.~Fulda, M.~Brehmer, and T.~Munzner.
\newblock {TimeLineCurator}: Interactive authoring of visual timelines from
  unstructured text.
\newblock {\em IEEE Transactions on Visualization and Computer Graphics},
  22(1):300--309, 2016.

\bibitem{gershon2001storytelling}
N.~Gershon and W.~Page.
\newblock What storytelling can do for information visualization.
\newblock {\em Communications of the ACM}, 44(8):31--37, 2001.

\bibitem{hu_vizml_2019}
K.~Hu, M.~A. Bakker, S.~Li, T.~Kraska, and C.~Hidalgo.
\newblock Vizml: A machine learning approach to visualization recommendation.
\newblock In {\em Proceedings of the CHI Conference on Human Factors in
  Computing Systems}, pp. 1--12, 2019.

\bibitem{hu_dive_2018}
K.~Hu, D.~Orghian, and C.~Hidalgo.
\newblock {DIVE}: A mixed-initiative system supporting integrated data
  exploration workflows.
\newblock In {\em Proceedings of the Workshop on Human-In-the-Loop Data
  Analytics}, pp. 1--7. ACM, 2018.

\bibitem{hullman2013deeper}
J.~Hullman, S.~Drucker, N.~H. Riche, B.~Lee, D.~Fisher, and E.~Adar.
\newblock A deeper understanding of sequence in narrative visualization.
\newblock {\em IEEE Transactions on Visualization and Computer Graphics},
  19(12):2406--2415, 2013.

\bibitem{kim2019datatoon}
N.~W. Kim, N.~Henry~Riche, B.~Bach, G.~Xu, M.~Brehmer, K.~Hinckley, M.~Pahud,
  H.~Xia, M.~J. McGuffin, and H.~Pfister.
\newblock Datatoon: Drawing dynamic network comics with pen+ touch interaction.
\newblock In {\em Proceedings of the 2019 CHI Conference on Human Factors in
  Computing Systems}, pp. 1--12, 2019.

\bibitem{kim2020gemini}
Y.~Kim and J.~Heer.
\newblock Gemini: A grammar and recommender system for animated transitions in
  statistical graphics.
\newblock {\em IEEE Transactions on Visualization and Computer Graphics},
  27(2):485--494, 2020.

\bibitem{kim2021gemini}
Y.~Kim and J.~Heer.
\newblock Gemini 2: Generating keyframe-oriented animated transitions between
  statistical graphics.
\newblock In {\em IEEE Visualization Conference (VIS)}, pp. 201--205, 2021.

\bibitem{kim2017graphscape}
Y.~Kim, K.~Wongsuphasawat, J.~Hullman, and J.~Heer.
\newblock Graphscape: A model for automated reasoning about visualization
  similarity and sequencing.
\newblock In {\em Proceedings of the CHI Conference on Human Factors in
  Computing Systems}, pp. 2628--2638, 2017.

\bibitem{kosara2013storytelling}
R.~Kosara and J.~Mackinlay.
\newblock Storytelling: The next step for visualization.
\newblock {\em Computer}, 46(5):44--50, 2013.

\bibitem{lan2021kineticharts}
X.~Lan, Y.~Shi, Y.~Wu, X.~Jiao, and N.~Cao.
\newblock Kineticharts: Augmenting affective expressiveness of charts in data
  stories with animation design.
\newblock {\em IEEE Transactions on Visualization and Computer Graphics},
  28(1):933--943, 2021.

\bibitem{lan2021understanding}
X.~Lan, X.~Xu, and N.~Cao.
\newblock Understanding narrative linearity for telling expressive
  time-oriented stories.
\newblock In {\em Proceedings of the CHI Conference on Human Factors in
  Computing Systems}, pp. 1--13, 2021.

\bibitem{lee2015more}
B.~Lee, N.~H. Riche, P.~Isenberg, and S.~Carpendale.
\newblock More than telling a story: Transforming data into visually shared
  stories.
\newblock {\em IEEE Computer Graphics and Applications}, 35(5):84--90, 2015.

\bibitem{liu2018data}
Z.~Liu, J.~Thompson, A.~Wilson, M.~Dontcheva, J.~Delorey, S.~Grigg, B.~Kerr,
  and J.~Stasko.
\newblock Data illustrator: Augmenting vector design tools with lazy data
  binding for expressive visualization authoring.
\newblock In {\em Proceedings of the CHI Conference on Human Factors in
  Computing Systems}, pp. 1--13, 2018.

\bibitem{luo_deepeye_2018}
Y.~Luo, X.~Qin, N.~Tang, and G.~Li.
\newblock Deepeye: Towards automatic data visualization.
\newblock In {\em IEEE International Conference on Data Engineering (ICDE)},
  pp. 101--112, 2018.

\bibitem{mackinlay_automating_1986}
J.~Mackinlay.
\newblock Automating the design of graphical presentations of relational
  information.
\newblock {\em ACM Transactions on Graphics}, 5(2):110--141, 1986.

\bibitem{mackinlay_show_2007}
J.~Mackinlay, P.~Hanrahan, and C.~Stolte.
\newblock Show me: Automatic presentation for visual analysis.
\newblock {\em IEEE Transactions on Visualization and Computer Graphics},
  13(6):1137--1144, 2007.

\bibitem{mckenna2017visual}
S.~McKenna, N.~Henry~Riche, B.~Lee, J.~Boy, and M.~Meyer.
\newblock Visual narrative flow: Exploring factors shaping data visualization
  story reading experiences.
\newblock In {\em Computer Graphics Forum}, pp. 377--387, 2017.

\bibitem{moritz2018formalizing}
D.~Moritz, C.~Wang, G.~L. Nelson, H.~Lin, A.~M. Smith, B.~Howe, and J.~Heer.
\newblock Formalizing visualization design knowledge as constraints: Actionable
  and extensible models in draco.
\newblock {\em IEEE Transactions on Visualization and Computer Graphics},
  25(1):438--448, 2018.

\bibitem{reimers-2019-sentence-bert}
N.~Reimers and I.~Gurevych.
\newblock Sentence-bert: Sentence embeddings using siamese bert-networks.
\newblock {\em arXiv preprint arXiv:1908.10084}, 2019.

\bibitem{ren2017chartaccent}
D.~Ren, M.~Brehmer, B.~Lee, T.~H{\"o}llerer, and E.~K. Choe.
\newblock Chartaccent: Annotation for data-driven storytelling.
\newblock In {\em IEEE Pacific Visualization Symposium (PacificVis)}, pp.
  230--239, 2017.

\bibitem{riche2018data}
N.~H. Riche, C.~Hurter, N.~Diakopoulos, and S.~Carpendale.
\newblock {\em Data-driven storytelling}.
\newblock CRC Press, 2018.

\bibitem{roth1994interactive}
S.~F. Roth, J.~Kolojejchick, J.~Mattis, and J.~Goldstein.
\newblock Interactive graphic design using automatic presentation knowledge.
\newblock In {\em Proceedings of the SIGCHI conference on Human Factors in
  Computing Systems}, pp. 112--117, 1994.

\bibitem{sarfraz_visualization_2002}
M.~Sarfraz.
\newblock Visualization of positive and convex data by a rational cubic spline
  interpolation.
\newblock {\em Information Sciences}, 146(1):239--254, 2002.

\bibitem{satyanarayan_authoring_2014}
A.~Satyanarayan and J.~Heer.
\newblock Authoring narrative visualizations with ellipsis: Authoring narrative
  visualizations with ellipsis.
\newblock {\em Computer Graphics Forum}, 33(3):361--370, 2014.

\bibitem{schlegel_interpolation_2012}
S.~Schlegel, N.~Korn, and G.~Scheuermann.
\newblock On the interpolation of data with normally distributed uncertainty
  for visualization.
\newblock {\em IEEE Transactions on Visualization and Computer Graphics},
  18(12):2305--2314, 2012.

\bibitem{segel2010narrative}
E.~Segel and J.~Heer.
\newblock Narrative visualization: Telling stories with data.
\newblock {\em IEEE Transactions on Visualization and Computer Graphics},
  16(6):1139--1148, 2010.

\bibitem{shi_task-oriented_2019}
D.~Shi, Y.~Shi, X.~Xu, N.~Chen, S.~Fu, H.~Wu, and N.~Cao.
\newblock Task-oriented optimal sequencing of visualization charts.
\newblock In {\em IEEE Visualization in Data Science (VDS)}, pp. 58--66, 2019.

\bibitem{shi2021autoclips}
D.~Shi, F.~Sun, X.~Xu, X.~Lan, D.~Gotz, and N.~Cao.
\newblock Autoclips: An automatic approach to video generation from data facts.
\newblock {\em Computer Graphics Forum}, 40(3):495--505, 2021.

\bibitem{shi_calliope_2021}
D.~Shi, X.~Xu, F.~Sun, Y.~Shi, and N.~Cao.
\newblock Calliope: Automatic visual data story generation from a spreadsheet.
\newblock {\em IEEE Transactions on Visualization and Computer Graphics},
  27(2):453--463, 2020.

\bibitem{shi2021communicating}
Y.~Shi, X.~Lan, J.~Li, Z.~Li, and N.~Cao.
\newblock Communicating with motion: A design space for animated visual
  narratives in data videos.
\newblock In {\em Proceedings of the CHI Conference on Human Factors in
  Computing Systems}, pp. 1--13, 2021.

\bibitem{shi2021understanding}
Y.~Shi, Z.~Li, L.~Xu, and N.~Cao.
\newblock Understanding the design space for animated narratives applied to
  illustrations.
\newblock In {\em Proceedings of the CHI Conference on Human Factors in
  Computing Systems}, pp. 1--6, 2021.

\bibitem{stolper2016emerging}
C.~D. Stolper, B.~Lee, N.~H. Riche, and J.~Stasko.
\newblock Emerging and recurring data-driven storytelling techniques: Analysis
  of a curated collection of recent stories.
\newblock {\em Microsoft Research}, 2016.

\bibitem{tang_extracting_2017}
B.~Tang, S.~Han, M.~L. Yiu, R.~Ding, and D.~Zhang.
\newblock Extracting top-k insights from multi-dimensional data.
\newblock In {\em Proceedings of the ACM International Conference on Management
  of Data}, pp. 1509--1524, 2017.

\bibitem{thevenaz2000image}
P.~Th{\'e}venaz, T.~Blu, and M.~Unser.
\newblock Image interpolation and resampling.
\newblock {\em Handbook of medical imaging, processing and analysis},
  1(1):393--420, 2000.

\bibitem{tong2018storytelling}
C.~Tong, R.~Roberts, R.~Borgo, S.~Walton, R.~S. Laramee, K.~Wegba, A.~Lu,
  Y.~Wang, H.~Qu, Q.~Luo, et~al.
\newblock Storytelling and visualization: An extended survey.
\newblock {\em Information}, 9(3):65--106, 2018.

\bibitem{ueng2005interpolation}
S.-K. Ueng and S.-C. Wang.
\newblock Interpolation and visualization for advected scalar fields.
\newblock In {\em VIS 05. IEEE Visualization}, pp. 615--622, 2005.

\bibitem{vartak_seedb_2015}
M.~Vartak, S.~Rahman, S.~Madden, A.~Parameswaran, and N.~Polyzotis.
\newblock Seedb: Efficient data-driven visualization recommendations to support
  visual analytics.
\newblock In {\em Proceedings of the VLDB Endowment International Conference on
  Very Large Data Bases}, pp. 2182--2193, 2015.

\bibitem{vaswani2017attention}
A.~Vaswani, N.~Shazeer, N.~Parmar, J.~Uszkoreit, L.~Jones, A.~N. Gomez,
  {\L}.~Kaiser, and I.~Polosukhin.
\newblock Attention is all you need.
\newblock {\em Advances in neural information processing systems},
  30(1):5998--6008, 2017.

\bibitem{wang_narvis_2019}
Q.~Wang, Z.~Li, S.~Fu, W.~Cui, and H.~Qu.
\newblock Narvis: Authoring narrative slideshows for introducing data
  visualization designs.
\newblock {\em IEEE Transactions on Visualization and Computer Graphics},
  25(1):779--788, 2018.

\bibitem{wang_datashot_2020}
Y.~Wang, Z.~Sun, H.~Zhang, W.~Cui, K.~Xu, X.~Ma, and D.~Zhang.
\newblock Datashot: Automatic generation of fact sheets from tabular data.
\newblock {\em IEEE Transactions on Visualization and Computer Graphics},
  26(1):895--905, 2019.

\bibitem{wang2018infonice}
Y.~Wang, H.~Zhang, H.~Huang, X.~Chen, Q.~Yin, Z.~Hou, D.~Zhang, Q.~Luo, and
  H.~Qu.
\newblock Infonice: Easy creation of information graphics.
\newblock In {\em Proceedings of the CHI Conference on Human Factors in
  Computing Systems}, pp. 1--12, 2018.

\bibitem{wills_autovis_2010}
G.~Wills and L.~Wilkinson.
\newblock Autovis: Automatic visualization.
\newblock {\em Information Visualization}, 9(1):47--69, 2010.

\bibitem{wittenbrink_ifs_1995}
C.~M. Wittenbrink.
\newblock Ifs fractal interpolation for 2d and 3d visualization.
\newblock In {\em IEEE Proceedings Visualization'95}, pp. 77--84, 1995.

\bibitem{wongsuphasawat_voyager_2016}
K.~Wongsuphasawat, D.~Moritz, A.~Anand, J.~Mackinlay, B.~Howe, and J.~Heer.
\newblock Voyager: Exploratory analysis via faceted browsing of visualization
  recommendations.
\newblock {\em IEEE Transactions on Visualization and Computer Graphics},
  22(1):649--658, 2015.

\bibitem{wongsuphasawat_voyager_2017}
K.~Wongsuphasawat, Z.~Qu, D.~Moritz, R.~Chang, F.~Ouk, A.~Anand, J.~Mackinlay,
  B.~Howe, and J.~Heer.
\newblock Voyager 2: Augmenting visual analysis with partial view
  specifications.
\newblock In {\em Proceedings of the CHI Conference on Human Factors in
  Computing Systems}, pp. 2648--2659, 2017.

\bibitem{yang2021design}
L.~Yang, X.~Xu, X.~Lan, Z.~Liu, S.~Guo, Y.~Shi, H.~Qu, and N.~Cao.
\newblock A design space for applying the freytag's pyramid structure to data
  stories.
\newblock {\em IEEE Transactions on Visualization and Computer Graphics},
  28(1):922--932, 2021.

\bibitem{zhao_chartstory_2021}
J.~Zhao, S.~Xu, S.~Chandrasegaran, C.~Bryan, F.~Du, A.~Mishra, X.~Qian, Y.~Li,
  and K.-L. Ma.
\newblock Chartstory: Automated partitioning, layout, and captioning of charts
  into comic-style narratives.
\newblock {\em arXiv preprint arXiv:2103.03996}, 2021.

\end{thebibliography}
